\documentclass{article}

\usepackage{preprint}
\usepackage{upgreek}				
\usepackage[utf8]{inputenc} 		
\usepackage[T1]{fontenc}    		
\usepackage[dvipsnames]{xcolor}	
\usepackage{hyperref}       		
\hypersetup{
    colorlinks = true,
    linkcolor = {NavyBlue},		
    anchorcolor = .,
    citecolor = {NavyBlue},		
    filecolor = .,				
    menucolor = .,
    runcolor = {cyan}, 
    urlcolor = {NavyBlue},		
    }
\usepackage{url}            
\usepackage{booktabs}       
\usepackage{amsfonts}       
\usepackage{nicefrac}       
\usepackage{microtype}      
\usepackage{fancyhdr}       
\usepackage{graphicx}       
\usepackage{setspace} 
\usepackage{soul} 
\usepackage{ragged2e} 
\usepackage[toc,page]{appendix}
\usepackage{longtable}
\usepackage{eso-pic}
\usepackage{tikz}
\usepackage[superscript]{cite}
\usepackage{array}
\usepackage[author={Corentin}]{pdfcomment}

\graphicspath{{./Figures/}}     

  \centering
\title{Chemical Control for the Morphogenesis of Conducting Polymer Dendrites in Water}

\author{
  Antoine Baron\textsuperscript{a,c}, 
  Corentin Scholaert\textsuperscript{a,c}, 
  David Gu\'{e}rin\textsuperscript{a},
  Yannick Coffinier\textsuperscript{a}, 
  Fabien Alibart\textsuperscript{b},
  \\
  \bf{S\'{e}bastien Pecqueur\textsuperscript{a}}\\
  \\
  a. IEMN, UMR 8520 \\
Univ. Lille, CNRS, Univ. Polytechnique Hauts-de-France\\
59000 Lille, France\\
  \\
  b. Laboratoire Nanotechnologies \& Nanosyst\`{e}mes (LN2)\\
CNRS, Universit\'{e} de Sherbrooke,
J1X0A5, Sherbrooke, Canada\\
  \\
  c. equally contributing first authors\\
  \\
  \texttt{sebastien.pecqueur@iemn.fr} \\
}

\begin{document}
\maketitle

\begin{abstract}
Conducting polymer dendrite (CPD) morphogenesis is an electrochemical process that unlocks the potential to implement \textit{in materio} evolving intelligence in electrical systems: As an electronic device experiences transient voltages in an open-space wet environment, electrically conductive structures physically change over time, programming the filtering properties of an interconnect as a non-linear analog device. 
Mimicking the self-preservation strategies of some sessile organisms, CPDs adapt their morphology to the environment they grow in. 
Either studied as an electrochemical experiment or as neuromorphic devices, the dependence of CPDs' electrical properties on the chemical nature of their environment is still unreported, despite the inter-dependence between the electrical properties of the electrogenerated material and the chemical composition of their growth medium. 
In this study, we report on the existing intrication between the nature and concentration of the electrolytes, electroactive compounds and co-solvents and the electrical and the electrochemical properties of CPDs in water. 
CPDs exhibit various chemical sensitivities in water: their morphology is highly dependent on the nature of the chemical resources available in their environment. 
The selection of these resources therefore critically influence morphogenesis.
Also, concentrations have different impacts on growth dynamics, conditioning the balance between thermodynamic and kinetic control on polymer electrosynthesis. 
By correlating the dependencies of these evolving objects with the availability of the chemical resources in an aqueous environment, this study proposes guidelines to tune the degree of evolution of electronic materials in water. 
Such hardware is envisioned to exploit the chemical complexity of real world environments as part of information processing technologies.
\end{abstract}

\raggedright

\newpage 

\justifying

\section{Introduction}

\begin{figure}
  \centering
  \includegraphics[width=1\columnwidth]{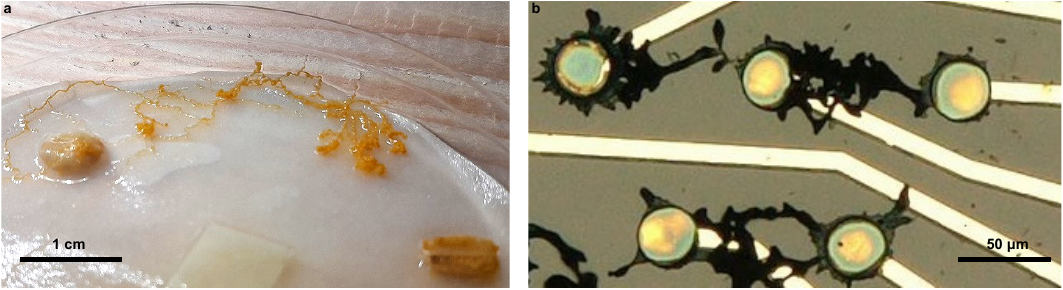}
  \caption{\textbf{Structural Analogy between Slime Mold and Conducting Polymer Dendrites in Water $\vert$ a,} \textit{physarum polycephalum} growing in a Petri dish between oat flakes in water. \textbf{b,} PEDOT:PSS growing on Parylene C between voltage-supplied electrodes in water.}
  \label{fig:fig1}
  \end{figure}

Human footprint on the ecosystem is associated with technological progress that promotes interactivity. 
With the deployment of Internet-of-Things technology and generative Artificial Intelligence models, more information is digitized, processed, shared and stored than ever before.\cite{Cisco2017,Reinsel2017} 
This surge in activity leads to a higher demand of electricity.\cite{deVries2023,EPRI2024,IEA2025} 
An important part of this increase is dedicated to the operation of data centers storing remotely-accessible information.\cite{Chen2025} 
A significant part of it also stems from manufacturing and retailing because of a globally decentralized economic model for end-consumer electronic trade.\cite{Pollard2021} 
In addition to the increasing demand for electricity, the maintenance of this interactivity requires perpetual consumption of electronic products: Outdated or damaged hardware must routinely be replaced.\cite{Solomon2000,Huang2019,Nunes2021} 
Due to low levels of recyclability and revalorization, this process significantly contributes to geological and environmental degradation because of the extraction of precious elements or biosourced materials.\cite{Vasan2014,Crawford2019,Dhar2020} 
Particularly in highly integrated systems, components are not meant to be replaced by the end-user unlike former detachable-block electronic items.\cite{Jones2014,Schischke2016,Hankammer2018} 
Also, at the opposite of software, conventional hardware does not upgrade by itself, so the design of a hardware system must be conceptualized \textit{a priori} to satisfy the broadest needs, featuring oversized resources.\cite{Salahuddin2018} 
In conventional electronic hardware, only electrons move and transience associated with materials (e.g. shorts or breaks) are undesirable and irrelevant for information processing.\\[3pt]
Evolutionary electronics is a broad concept,\cite{Zebulum2001} which proposes to exploit the transience of circuits with evolvable hardware as part of information processing.\cite{Yao1999,Haddow2011} 
Conventional computers do not customize themselves according to how we use them. 
Evolutionary electronics allow information to drive modifications in the physical structure to support vertical scaling, adapting the resources to the needs of a task. 
It offers to solve technical issues, such as functional block failures through self-repair, improved management of limited energy supplies and specific allocation of computing and memory resources based on user experience. 
This paradigm draws inspiration from nature, and the necessity of some biological systems to physically adapt their structure as an information processing resource: from the brain topological plasticity to the networking behavior of sessile organisms.\cite{Stepanyants2002,Chaiwanon2016} 
Some organisms, such as slime mold (Fig.\ref{fig:fig1}.a) have genuine abilities to solve classification tasks and have been proposed as low-resource computing materials in water.\cite{Tero2010,Adamatzky2010}\\[3pt]
Although computers have historically developed around hardware, the substrate for computation does not always have to be solid: Liquid computers take advantage of fluid dynamics to perform computing.\cite{Adamatzky2019} 
Drawing inspiration from the operation of the brain and the nervous system, which consist of wetware,\cite{Maass2001} neuromorphic devices functioning in aqueous environments have started to emerge,\cite{Jung2023} and liquid-based devices have been shown to implement physical reservoirs in the framework of reservoir computing.\cite{Sato2023} 
Aside from computation, information storage in liquid samples is also being explored for data archiving.\cite{Bornholt2016,Akram2018,Ceze2019} 
Although accessing the data requires a sequencer, being able to write information on macromolecules in water is advantageous for a number of reasons, such as high density information storage at small scale in a three dimensional volume, ease of copying/sharing and physical robustness of the information carrier over long periods of time without energy supply. 
Thus, and because water is a universal substrate for biologically intelligent organisms, there is a necessity to investigate on water-based machines able to compute and store information and communicate with other organisms in the same medium.\\[3pt]
In this scope, conducting polymer morphogenesis exploits voltage-driven electropolymerization to promote the growth of conducting materials on electrodes immersed in an ion-containing environment (Fig.\ref{fig:fig1}.b). 
With a transient voltage, conducting polymer dendrites (CPDs) grow as dark fibrous structures and exhibit high versatility in aqueous environments.\cite{Janzakova2021b} 
They resemble the ramified structures found in nature. 
They are highly disordered, but their structural features are globally controlled by voltage. 
The mechanism ruling their charge coupling with ions in their medium and their structural complexity provide many advantages to use CPD networks as a computing material.\cite{AkaiKasaya2020,Petrauskas2021,Janzakova2023,Scholaert2024}
In water, a train of low voltage spikes supplied between two electrodes can condition their growth to connect specific contacts.\cite{Janzakova2021a}. 
When stimulated with a dynamical voltage signal, the structural asymmetry is a relevant property for transient pattern classification.\cite{Scholaert2022,Scholaert2024} 
Due to the electrochemical nature of the admittance that connects them, the tuning of their morphology greatly conditions information transport by modifying their filtering properties.\cite{Baron2024a,Baron2024,Baron2025}\\[3pt]
Up to now, CPDs have been studied under only a few very specific conditions (Table \ref{table:tab1}). 
Their growth has shown to depend on the nature of the monomer, the redox-active counter agent, the electrolyte salt and the solvent, but quantitative correlations have not been performed. 
Also, growth was mostly studied in acetonitrile, a volatile organic solvent. 
However, the first tests reported in water in a comparative study with acetonitrile show that the morphology and behavior of the obtained dendrites can be very different.\cite{Koizumi2018} 
This study reports on the influence of the chemical composition of an aqueous electrolyte on the morphogenesis of CPDs.\\

\begin{table}[!htbp]
\centering
\caption{Growth conditions studied in the literature and in the present work}
\begin{tabular}{*5c}
  \toprule
   Monomer & Redox Agent & Electrolyte & Solvent & Reference \\ 
  \midrule
  500~mM pyrrole & & 125~mM & aqueous & Dang \textit{et al.}, 2014\cite{Dang2014} \\
  135~mM EDOT & & 20~mM & aqueous & Dang \textit{et al.}, 2014\cite{Dang2014} \\
  50~mM EDOT & 5~mM BQ & 1~mM Bu\textsubscript{4}NClO\textsubscript{4} & MeCN & Koizumi, Ohira \textit{et al.}\cite{Koizumi2016,Ohira2017} \\
  50~mM EDOT-CH\textsubscript{3} & 5~mM BQ & 1~mM Bu\textsubscript{4}NClO\textsubscript{4} & MeCN & Koizumi \textit{et al.}, 2016\cite{Koizumi2016} \\  
  50~mM EDOT-C\textsubscript{10}H\textsubscript{21} & 5~mM BQ & 1~mM Bu\textsubscript{4}NClO\textsubscript{4} & MeCN & Koizumi \textit{et al.}, 2016\cite{Koizumi2016} \\    50~mM EDOT & \multicolumn{2}{c}{--------- 1~mM H\textsubscript{2}PtCl\textsubscript{6} ---------} & MeCN & Koizumi \textit{et al.}, 2018\cite{Koizumi2018} \\
  50~mM EDOT & \multicolumn{2}{c}{-------------- AgBF\textsubscript{4} --------------} & MeCN & Koizumi \textit{et al.}, 2018\cite{Koizumi2018} \\
  10~mM EDOT & & 1~g/L NaPSS & \textbf{H\textsubscript{2}O} & Koizumi \textit{et al.}, 2018\cite{Koizumi2018} \\
  10~mM EDOT & & NaTs & \textbf{H\textsubscript{2}O} & Koizumi \textit{et al.}, 2018\cite{Koizumi2018} \\
  150~mM EDOT & 15~mM BQ & 3~mM Bu\textsubscript{4}NClO\textsubscript{4} & MeCN & Watanabe \textit{et al.}, 2018\cite{Watanabe2018} \\
  150~mM EDOT & 15~mM BQ & 3~mM  Bu\textsubscript{4}NBF\textsubscript{4} & MeCN & Watanabe \textit{et al.}, 2018\cite{Watanabe2018} \\
  150~mM EDOT & 15~mM BQ & 3~mM  Bu\textsubscript{4}NPF\textsubscript{6} & MeCN & Watanabe \textit{et al.}, 2018\cite{Watanabe2018} \\
  150~mM EDOT-CH\textsubscript{3} & 15~mM BQ & 3~mM Bu\textsubscript{4}NClO\textsubscript{4} & MeCN & Watanabe \textit{et al.}, 2018\cite{Watanabe2018} \\
  150~mM EDOT-C\textsubscript{10}H\textsubscript{21} & 15~mM BQ & 3~mM Bu\textsubscript{4}NClO\textsubscript{4} & MeCN & Watanabe \textit{et al.}, 2018\cite{Watanabe2018} \\
  150~mM EDOT-CH\textsubscript{2}Cl & 15~mM BQ & 3~mM Bu\textsubscript{4}NClO\textsubscript{4} & MeCN & Watanabe \textit{et al.}, 2018\cite{Watanabe2018} \\
  150~mM bithiophene & 15~mM BQ & 3~mM Bu\textsubscript{4}NClO\textsubscript{4} & MeCN & Watanabe \textit{et al.}, 2018\cite{Watanabe2018} \\
  50~mM EDOT & 10~mM BQ & \multicolumn{2}{c}{----------- [Deme]BF\textsubscript{4} -----------} & Chen \textit{et al.}, 2023\cite{Chen2023} \\ 
  50~mM EDOT & 10~mM BQ & \multicolumn{2}{c}{----------- [Emim]BF\textsubscript{4} -----------} & Chen \textit{et al.}, 2023\cite{Chen2023} \\
  50~mM EDOT & 10~mM BQ & \multicolumn{2}{c}{----------- [Deme]NTf\textsubscript{2} -----------} & Chen \textit{et al.}, 2023\cite{Chen2023} \\ 
  50~mM EDOT & & 1~mM Bu\textsubscript{4}NClO\textsubscript{4} & MeCN & Eickenscheidt \textit{et al.}, 2019\cite{Eickenscheidt2019} \\
  50~mM EDOT & & 1~mM Bu\textsubscript{4}NPF\textsubscript{6} & MeCN & Cucchi, Petrauskas \textit{et al.}\cite{Cucchi2021,Cucchi2021a,Petrauskas2021} \\
  135~mM EDOT & & 20~mM "PSS" & \textbf{H\textsubscript{2}O}/MeCN & Akai Kasaya, Hagiwara,\\
  & & & & Watanabe \textit{et al.}\cite{AkaiKasaya2020,Hagiwara2021,Hagiwara2023,Watanabe2024} \\
  10~mM EDOT & 10~mM BQ & 1~mM NaPSS & \textbf{H\textsubscript{2}O} & Janzakova, Scholaert, \\
  & & & & Baron \textit{et al.}\cite{Janzakova2021,Janzakova2021a,Janzakova2021b,Janzakova2023,Scholaert2022,Baron2024,Baron2024a,Scholaert2024,Baron2025}   \\
  \cmidrule{1-5}
  10~mM EDOT & 10~mM BQ & $\upmu$M-mM NaPSS & \textbf{H\textsubscript{2}O} & \textit{This work} (Fig.\ref{fig:fig2}) \\
  10~mM EDOT & 10~mM BQ & 1~mM EmimOTf & \textbf{H\textsubscript{2}O} & \textit{This work} (Fig.\ref{fig:fig3}) \\
  \multicolumn{2}{c}{----- 10~mM \textit{x}EDOT+\textit{(1-x)}BQ -----} & 1~mM NaPSS & \textbf{H\textsubscript{2}O} & \textit{This work} (Fig.\ref{fig:fig4}) \\
  10~mM EDOT & 10~mM BQ & 1~mM NaPSS & \textbf{H\textsubscript{2}O}/glycerol & \textit{This work} (Fig.\ref{fig:fig5}) \\
  2.5-10~mM EDOT & 10~mM BQ & 1~mM NaPSS & \textbf{H\textsubscript{2}O} & \textit{This work} (Fig.\ref{fig:fig6}) \\
  10~mM EDOT(g\textsubscript{4})\textsubscript{x} & 10~mM BQ & 1~mM NaPSS & \textbf{H\textsubscript{2}O} & \textit{This work} (Fig.\ref{fig:fig7}) \\
  10~mM EDOT(RSO\textsubscript{3}Na)\textsubscript{x} & 10~mM BQ & & \textbf{H\textsubscript{2}O} & \textit{This work} (Fig.\ref{fig:fig8}) \\
  \bottomrule
    \label{table:tab1}
\end{tabular}
\end{table}
EDOT: 3,4-ethylenedioxythiophene\\
EDOT-CH\textsubscript{4}: 2-methyl-2,3-dihydrothieno[3,4-b][1,4]dioxine\\
EDOT-C\textsubscript{10}H\textsubscript{21}: 2-decyl-2,3-dihydrothieno[3,4-b][1,4]dioxine\\
EDOT-CH\textsubscript{2}Cl: 2-chloromethyl-2,3-dihydrothieno[3,4-b][1,4]dioxine\\
BQ: parabenzoquinone\\
EDOTg\textsubscript{4}: 2-(2,5,8,11-tetraoxadodecyl)-2,3-dihydro-thieno-
[3,4-b][1,4]dioxine\\
EDOT(RSO\textsubscript{3}Na): Sodium 4-(2,3-dihydrothieno[3,4-b][1,4]-
dioxin-2-yl-methoxy)-1-butanesulfonate\\
Bu\textsubscript{4}NClO\textsubscript{4}: tetrabutylammonium perchlorate\\
NaPSS: sodium polystyrene sulfonate\\
NaTs: sodium tosylate\\
Bu\textsubscript{4}NBF\textsubscript{4}: tetrabutylammonium tetrafluoroborate\\
Bu\textsubscript{4}NPF\textsubscript{6}: tetrabutylammonium hexafluorophosphate\\
{[}Deme{]}BF\textsubscript{4}: diethylmethyl(2-methoxyethyl)ammonium tetrafluoroborate\\
{[}Emim{]}BF\textsubscript{4}: 1-ethyl-3-methylimidazolium tetrafluoroborate\\
{[}Emim{]}NTf\textsubscript{2}: 1-ethyl-3-methylimidazolium bis(trifluoromethanesulfonyl)imide\\
{[}Emim{]}OTf: 1-ethyl-3-methylimidazolium trifluoromethanesulfonate\\
MeCN: acetonitrile\\

\newpage

\section{Experimental}

\subsection{Materials and Methods}

All purchased chemicals were used without further purification and were exposed to air, ambient light and room temperature. 
3,4-Ethylenedioxythiophene (EDOT), glycerol, sodium polystyrene sulfonate (NaPSS, 70~kDa) and parabenzoquinone (BQ) were purchased from Sigma Aldrich. 
1-Ethyl-3-methylimidazolium trifluoromethanesulfonate (emimOTf) was purchased from Solvionic. Sodium 4-(2,3-dihydrothieno[3,4-b][1,4]-
dioxin-2-yl-methoxy)-1-butanesulfonate (EDOT(RSO\textsubscript{3}Na) or EDOT-S)\cite{Cutler2005} was purchased from BLD Pharm. 2-(2,5,8,11-tetraoxadodecyl)-2,3-dihydro-thieno-
[3,4-b][1,4]dioxine (EDOTg\textsubscript{4}) was synthesized as reported by Ghazal \textit{et al.}\cite{Ghazal2023} CPD growth was performed on 25~$\upmu$m diameter gold wires purchased from Goodfellow.\\[3pt]

\subsection{Electrical Characterization}

The electrical conductance of the dendrites was estimated by performing current-voltage (I-V) measurements with the help of an Agilent B1500A Semiconductor Analyzer coupled with a B2201A Switching Matrix.\\[3pt]
Impedance measurements were performed using a Solartron Analytical (Ametek) impedance analyzer, with the measurement taken between the shorted ends of the CPDs and a 25~$\upmu$m diameter silver wire serving as a counter electrode.\\[3pt]
The transfer characteristic of the EDOT-S OECT was obtained by applying a voltage bias of 100~mV between the source and drain electrodes, the source being grounded, and a voltage varying between {\textminus}500~mV and +500~mV to the gate relative to the source using a hold time of one second and a delay of 100~ms. 
Voltage application was performed and current measurements were obtained using the SMU channels of a B1500A Semiconductor Device Analyzer.\\[3pt]
Unless otherwise specified, CPDs were grown using a square wave voltage signal centered at 0~V, alternating between {\textminus}4~V and +4~V at a frequency of 80~Hz and a 50\% duty cycle, generated by a 50MS/s Dual-Channel Arbitrary Waveform Generator from Tabor Electronics.\\[3pt]

\subsection{Imaging}

The optical microscope pictures presented throughout this paper were extracted from video footage captured using a VGA CCD color Camera from HITACHI Kokusai Electric Inc, on dendrites suspended on the gold wires in the electrolyte solution used for the growth.\\[3pt]
Scanning Electron Microcope (SEM) pictures were subsequently taken using the InLens detector of a Zeiss Ultra 55, after careful removal of the gold wires from the electrolyte droplet, taped on a silicon substrate. \\[3pt]

\newpage 

\section{Results}

\subsection{Media Salinity and Morphogenesis of Conducting Polymer Dendrites}

\begin{figure}
	\centering
	\includegraphics[width=1\columnwidth]{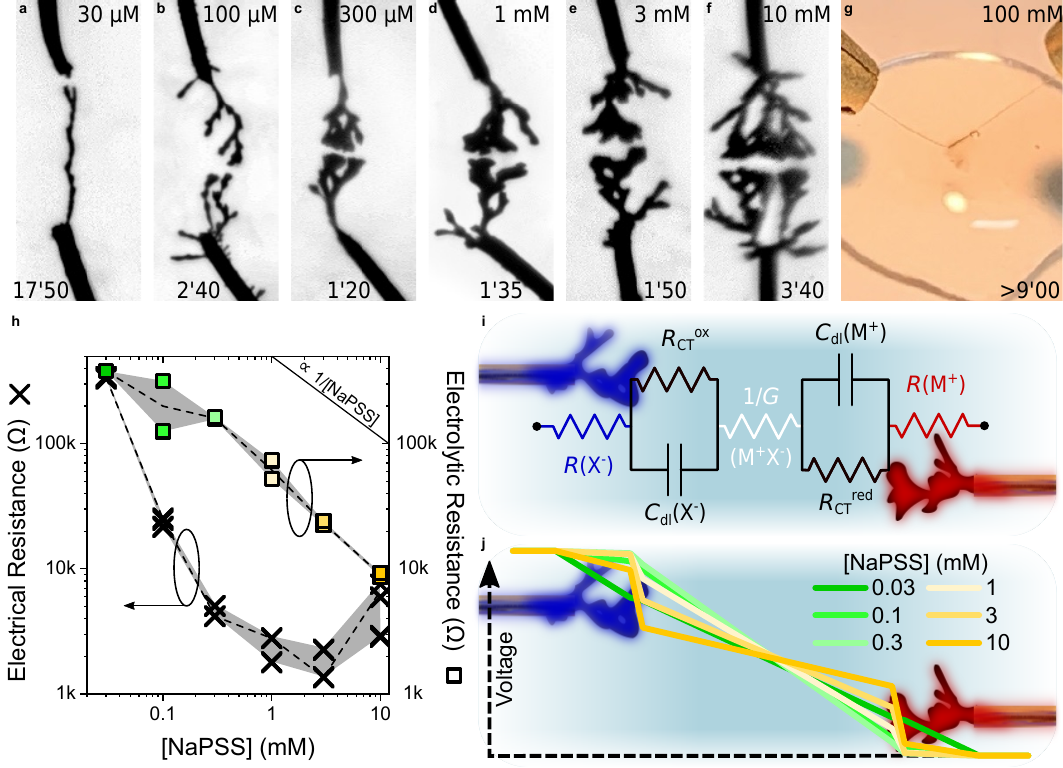}
	\caption{\textbf{Influence of the Concentration of NaPSS on CPD Morphogenesis in Water $\vert$ a-f,} Optical pictures of CPD growths (4~V\textsubscript{p}, 0~V\textsubscript{off}, 80~Hz, 50\%\textsubscript{dc}, 10~mM EDOT and 10~mM BQ) in aqueous media with different NaPSS concentrations (values displayed at the top of each picture). Each image was taken a few seconds before CPDs merged (applied-voltage duration displayed at the bottom of each picture). \textbf{g,} Photograph of the electrochemical setup under the same experimental conditions as previously described, but with 100~mM of NaPSS. Despite the absence of CPD between the wires, the presence of an electrogenerated material is visible. \textbf{h,} Dependence of the electrical resistance (through each CPD displayed in Fig.\ref{fig:fig3}.a-f after they merged) and the electrolytic resistance (evaluated by impedancemetry at 10~kHz with a silver wire) on the concentration of NaPSS in the growth media. \textbf{i,} 7-Element equivalent circuit highlighting the contribution of [NaPSS] on the ohmic resistance of the medium and the CPDs as both an electrolyte and a doping counter-ion carrier. \textbf{j,} Simulated voltage drop according to the equivalent circuit presented in Fig.\ref{fig:fig3}.i, with R\textsubscript{CT}\textsuperscript{ox}~=~R\textsubscript{CT}\textsuperscript{red}~=~10~k$\Omega$, \textit{R}(X\textsuperscript{-})~=~\textit{R}(M\textsuperscript{+}) defined as half the experimental value of the CPD electrical resistance once merged and 1/\textit{G} as the experimental value of the electrolytic resistance measured with a silver wire.}
	\label{fig:fig2}
\end{figure}

While direct-current (DC) electropolymerization is usually voltage activated using a potentiostat and performed in 100~mM concentrated electrolytes, CPD growth is electric-field activated using an alternating-current (AC) waveform induced by a voltage generator.\cite{Koizumi2016} 
Regardless of the nature of the solvents or electrolytes employed, the growth of PEDOT-based CPDs is usually performed under mild saline conditions to promote electric-field activation in a medium behaving more as a dielectric than as an ion conductor (Table \ref{table:tab1}). 
Controlling the balance between the dielectric and conductive properties of water is thus expected to significantly influence CPD growth: the concentration of salt shall therefore be of major importance. 
In water, PSS is a polyanion extensively used in the literature because (1) it stabilizes the p-doping of PEDOT as an oxidized conducting polymer, (2) it promotes cation conduction in a solid PEDOT:PSS blend as an organic mixed ionic-electronic conductor (OMIEC)\cite{Paulsen2020,Tropp2023,Wang2023,Wu2023,Gkoupidenis2024} and (3) as an emulsifier, it stabilizes the hydrophobic PEDOT in water. 
Akai-Kasaya and coworkers reported that among different anions tested in a water/acetonitrile blend, PSS provided the highest degree of controllability and reproducibility, yielding thin and wire-like growths.\cite{AkaiKasaya2020} 
However, the numerous roles played by this anion during electropolymerization hinder a full understanding of its involvement during CPD growth, challenging the identification of an issue when growth is not observed.\\[3pt]
The morphological influence of the concentration of NaPSS can be observed in Fig.\ref{fig:fig2}.a-f, which presents pictures of CPDs taken seconds prior completion at different concentrations of salt, while all other parameters were fixed---nature of the wires and of the electrolyte and parameters of the voltage waveform. 
Fiber growth was found to happen for concentrations ranging from 30~$\upmu$M to 10~mM. 
Higher and lower concentrations of salt did not lead to dendritic growth experimentally, although it can be noted that, at 100~mM (Fig.\ref{fig:fig2}.g), a dark veil appeared around the electrodes and some material was deposited onto the gold wires, but no dendrite grew and bubbles were formed due to water electrolysis. 
CPD can grow in ionic liquids,\cite{Chen2023} suggesting that the limitation observed at high NaPSS concentrations does not result from screening of the applied electric field by an excessive ionic density. 
Several hypotheses can be raised to explain this phenomenon: At these high concentrations, PEDOT might be too well dispersed in the medium, preventing CPDs from sticking on the gold electrodes as the anion behaves as a surfactant. 
As NaPSS also operates as a doping counter-anion, excessive doping of PEDOT may result in highly charged particles, which in turn favors their suspension in the medium. 
These observations raise questions about the possibility to perform CPD growth in physiological environments, where salt concentration reaches such ranges, although this issue could be circumvented by leveraging the metabolic activity of living organisms to perform \textit{in vivo} polymerization.\cite{Tommasini2022, Stavrinidou2017, Dufil2020, Priyadarshini2023, Strakosas2023}\\
The morphology of the electrogenerated CPDs appeared to change with the concentration of NaPSS in the electrolyte. 
Lower concentrations resulted in more linear structures, whereas higher concentrations led to more voluminous and branched systems, with the appearance of the fibers gradually evolving with concentration. 
Growth duration was also significantly affected by the concentration of NaPSS available in the medium, with a minimum of 100 seconds found to bridge a 240~$\upmu$m gap with 1~mM NaPSS in water (Fig.\ref{fig:fig2}.e), while at 30~$\upmu$M, the CPD took significantly longer to grow---it even required narrowing the gap at the nucleation phase to initiate growth on the gold surface.\\[3pt]
The electrical properties of the electrogenerated objects as well as those of the medium were recorded at the end of completion. 
Fig.\ref{fig:fig2}.h presents the resistance of the dendrites calculated through I-V measurements as well as the resistance of the electrolyte as estimated by impedancemetry. 
As expected, the conductivity of the solution increases quasi-linearly with salt concentration.  
More interestingly, the electrical resistance of the polymer fibers appears to decrease as the concentration of NaPSS increases up to 3~mM, before slightly increasing at 10~mM. 
Two effects might contribute. 
First, as salt concentration increases, the dendrites tend to become thicker, thus increasing their conductance. 
Second, as NaPSS acts as the dopant in PEDOT:PSS, increasing its concentration in the growth solution might increase the conductivity of the polymer. 
As estimating the volume of these fractal objects remains a challenge, it is difficult to estimate their conductivity, and therefore to decorrelate the influence of each effect. 
Yet, these results show that salt concentration is a parameter that influences the time it takes for the fibers to form, their morphology as well as the conductance of these objects.\\[3pt]
The influence of salt concentration on CPD formation can be appreciated through the simulations of voltage drop in the system during growth presented in Fig.\ref{fig:fig2}.i-j. 
A low-complexity equivalent circuit was used to model the system, which consisted in a hole transport resistance \textit{R} for each end of the CPD, a double layer capacitance \textit{C}\textsubscript{dl} in parallel with a charge-transfer resistance (\textit{R}\textsubscript{CT}\textsuperscript{ox} and \textit{R}\textsubscript{CT}\textsuperscript{red}) that models the redox reactions at the two polymer/electrolyte interfaces, and the electrolytic resistance 1/\textit{G} governed by ion transport in the medium (Fig.\ref{fig:fig2}.i). 
As a first-level approximation, it was assumed that the resistance of both uncompleted CPDs was independent of the applied voltage and equaled half the measured electrical resistance displayed in Fig.\ref{fig:fig2}.h. 
It was also considered that the electrolytic resistance 1/\textit{G}, which was limited by the ion drift between two CPDs, was also representative of the impedance between the electrodes and the silver wire used for the impedance measurement. 
Additionally, both \textit{R}\textsubscript{CT} were fixed to 10~k$\Omega$ and both \textit{C}\textsubscript{dl} were considered ideal Debye capacitors. 
As presented in Fig.\ref{fig:fig2}.j, the results show that the concentration of NaPSS influences the voltage drops in the system through the modulation of both the electrolytic resistance and the electrical resistance of the dendrite. 
For low salt concentrations, the voltage drop across the fibers cannot be neglected because of the low conductance of CPDs. 
As a consequence, the potential is not uniformly distributed along the fibers. 
The voltage drop at the electrode/electrolyte interfaces increases with the concentration of NaPSS, therefore driving electropolymerization at the tip of the electrodes.
Despite the influence of salt concentration on the morphology of CPDs, these results confirm that two antagonistic effects - both governed by the presence of salt in the medium - interplay to control growth kinetics through their influence on the voltage drop: insufficient salt leads to poorly conductive CPDs, while excessively high salt concentrations result in a overly conductive electrolytes.\\[3pt]

\begin{figure}
  \centering
  \includegraphics[width=1\columnwidth]{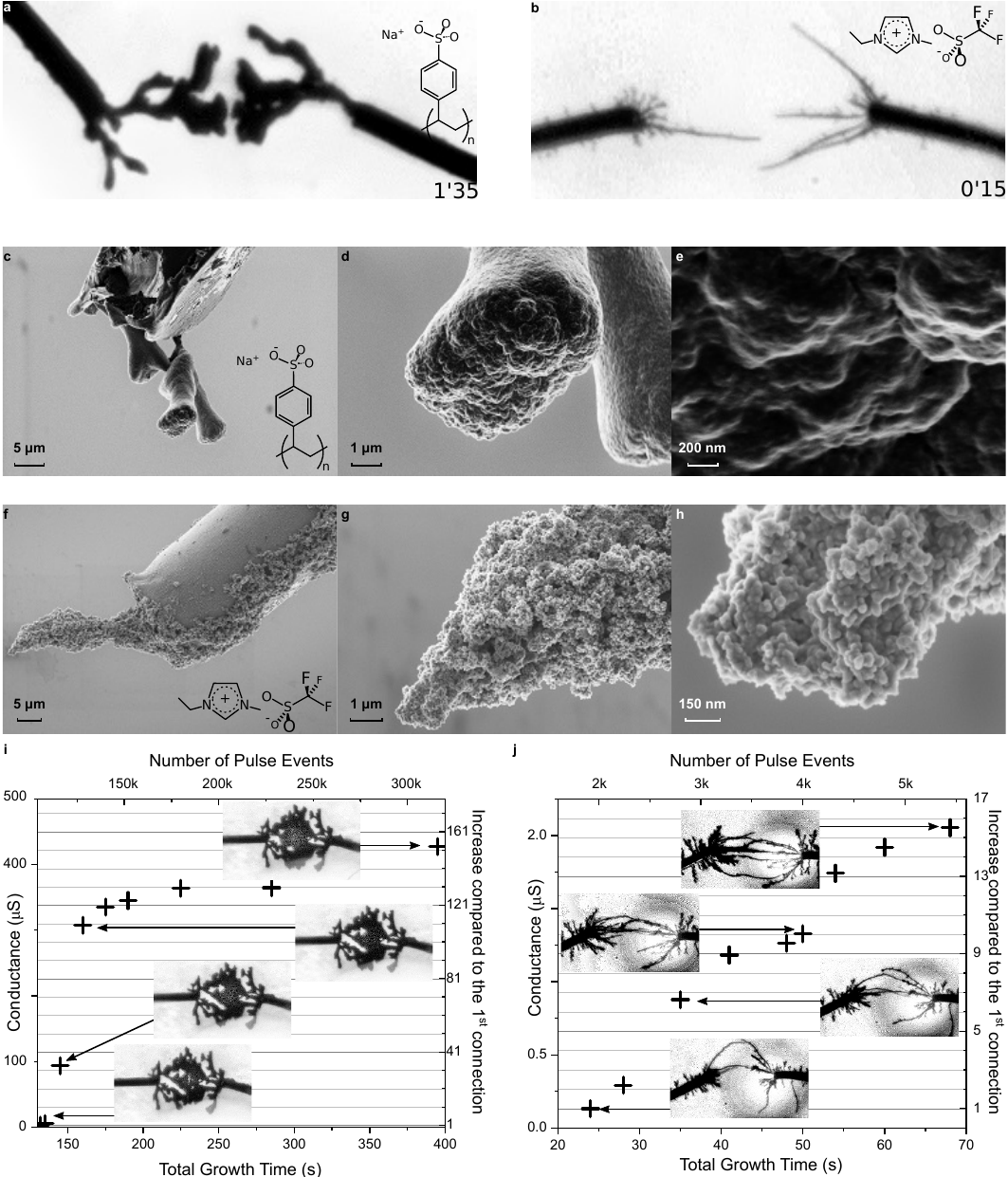}
  \caption{\textbf{Influence of the Nature of a Salt on a CPD Topology in Water $\vert$ a,} Optical picture of a CPD growth at 4~V\textsubscript{p}, 0~V\textsubscript{off}, 80~Hz, 50\%\textsubscript{dc} in water with 10~mM EDOT and 10~BQ at 1~mM NaPSS afer 95 seconds of applied voltage. \textbf{b,} Optical picture of a CPD growth n exacly the same conditions as Fig.\ref{fig:fig3}.a, but using EmimOTf instead of NaPSS as an electrolyte and after only 15 seconds of applied voltage. \textbf{c-e,} Scanning Electron Microscope (SEM) images of a CPD growth in the same conditions as displayed in \ref{fig:fig3}.a in the presence of NaPSS as an electrolyte. \textbf{f-h,} SEM images of a CPD growth in the same conditions as displayed in \ref{fig:fig3}.b in the presence of EmimOTf as an electrolyte. \textbf{i,j,} Filamentary switching: \textbf{i} shows the evolution of conductance for PEDOT:PSS CPDs over time, starting after an initial connection is completed. Each sample approximately corresponds to the connection of a new branch. \textbf{j} shows the same experiment for PEDOT:OTf CPDs. Insets are black and white optical microscope pictures of the corresponding CPDs at different times.}
  \label{fig:fig3}
  \end{figure}

It has recently been established that CPD can grow in ionic liquids (ILs)\cite{Chen2023} and that they exhibit different morphologies depending on the diffusion coefficients of both ions present in solution. 
However, ILs are often expensive, and highly viscous, which can considerably lower the growth speed, even in the presence of high ion concentrations. 
A high viscosity also prevents the formation of electrohydrodynamic instabilities that have been suspected to be involved in the formation of dendritic deposits.\cite{Fleury_1993} 
Whether the observed difference of morphology comes from the difference in viscosity of the medium or the chemistry of the ions is therefore still an open question. 
Based on the data in Table \ref{table:tab1}.a, CPDs are only observed to grow under conditions favoring a diffusion-limited current, i.e either high viscosity or low ionic concentration, which does not rule out the possibility of growing CPDs with different counter ions than PSS, as long as they remain diluted and their mobility compatible with the viscosity of the medium. 
Since water has a low viscosity, the mobility in water is inherently high and in order to reach a diffusion-limited regime, the concentration of the anions is usually kept low.
In contrast, when the concentration is too high, it leads to a reaction-limited growth current producing compact morphologies such as in the case of Fig.\ref{fig:fig2}.g.
As EmimOTf is an IL miscible in water and given that it was previously used to successfully form a PEDOT:OTf polymer,\cite{Gueye2016} it raises the question whether triflate ions (OTf\textsuperscript{-}) can also support the growth of CPDs in water, and if the resulting CPDs are different depending on the nature of the counter ion. 
To answer these questions, a comparison was made between PEDOT:PSS and PEDOT:OTf CPDs. Both electrolytes were introduced at 1~mM as it was previously established to be the most optimal concentration for fast growth, with 10~mM EDOT and 10~mM BQ in water.
The applied growth signal was identical in both cases, making the nature of the doping counter anion the only difference between the two experiments. 
It was observed that the branches are thicker in the case of NaPSS and that they will not grow on the sides of the wire. 
Using EmimOTf however will result in short branches growing on the sides and thinner, more linear CPDs. 
The growth itself is extremely fast in the case of PEDOT:OTf, taking only around 15 seconds to complete, more than a minute for PEDOT:PSS. 
It is hypothesized that the observed differences mainly come from a difference in ionic mobility, for a number of reasons. 
Indeed, in the case of the NaPSS salt, PSS is a polyanion with high molecular weight (70~kDa used in the present study), while triflate ions only carry a single charge. 
Since the growth medium is water, the anion drift is hindered by friction, which increases with molecular size. 
Moreover, although PSS bears one negative charge per monomer unit in its chemical structure, its effective charge is often significantly reduced by counterion condensation,\cite{Adamczyk2009} making it a massive molecule with a lower effective charge density.
Therefore, the faster growth dynamics of triflate ions could be explained by their greater mobility. 
PEDOT:OTf CPDs growth follows the paths of least resistance dictated by the field between the two wires, leaving less time for diffusive and electrohydrodynamic effects to alter the ions trajectory. 
The high mobility also implies that ions will cross greater distances for the same amount of time, leading to OTf\textsuperscript{-} ions reaching the sides of the wire on a wider range than PSS in the same conditions. 
Interestingly, side branches grow from the side in the direction of the electric field, perpendicularly to the wire surface, rather than following the direction of the wire, indicating that they are more of an effect of the electric field rather than diffusion. 
The resulting CPDs have been compared in a dried state by scanning electron microscopy. 
Fig.\ref{fig:fig3}.c–e show the same PEDOT:PSS CPD observed at different magnifications. 
The images confirm that deposition on the sides of the wire is limited, with most of the growth occurring at the tip in the form of a fractal CPD object. 
The sides of the dendritic structure appear notably smooth, whereas the tip displays multiple nucleation sites concentrated within the frontal surface of the CPD tip, which exhibits a significantly higher roughness. 
An important detail to note is that this roughness is not the result of the CPD breaking during sample handling. 
In addition to this microscopic roughness, the frontal surface is not flat: a depression is observed at the center, with more protruding regions on the sides. 
This suggests that, after a critical growth time, the surface splits, and one branch becomes favored over the other depending on whether it follows the path of least resistance. 
Eventually, the "dead" branch appears to smoothen in surface, possibly due to the combined effect of water absorption leading to swelling and the surfactant properties of the PSS ions, which may promote surface rearrangement and smoothing.\\[3pt]
PEDOT:OTf CPDs were also observed under the SEM, as shown in Fig.\ref{fig:fig3}.f–h. 
Although the CPD is more fragile and can break due to the surface tension of the water droplet, the morphology of the PEDOT:OTf coating visible along the side of the wire corresponds to the remains of the CPD at the tip. 
Made from the same monomer in the same solvent, the roughness is significantly higher, with no apparent difference between the surface and the core of the CPD. 
The material appears to be composed of grains smaller than 50~nm, seemingly without any binder as seen with PSS. 
Although this high roughness provides multiple potential nucleation sites, the main branches rarely split. 
This difference in growth modes can be employed in various applications, such as filamentary switching, which involves modulating the conductance of a connection through the formation and rupture of conductive filaments.\cite{LaBarbera2015,Zhang2022} 
After the initial completion of the dendritic connection, the same growth signal is reapplied and I–V measurements are performed at different times to evaluate the possible irreversible modulation of conductance after connection. 
In the case of PEDOT:PSS dendrites, sudden conductance jumps as high as 200~$\upmu$S can occur, resulting from the growth of inner branches at the junction.
The PEDOT:PSS large branches connecting together results in a higher electronic current increase because of the sudden gain in active surface area, and gradually, the connection is reinforced by branches merging together around the junction, eventually forming a single branch if given enough time.
On the other hand, the PEDOT:OTf CPDs show a more linear increase in conductance due to the continuous connection of new branches between the two electrodes. 
Side branches, considered "dead" before the connection, are developing again and contribute to strengthen the connection either ionically or electronically when connecting to the nearest branches, possibly to one belonging to the same side of the system. 
The conductance of PEDOT:PSS, highly dependent on the active surface, can reach two orders of magnitude higher than PEDOT:OTf. 
By changing the nature of only the counterion, not only is it possible to get different conductivities for the material, but the modulation of conductance and its dynamics can also be modified to better suit specific applications.

\subsection{Electrochemical Control of the Growth Velocity by Charge-Transfer Balance}

\begin{figure}
  \centering
  \includegraphics[width=\columnwidth]{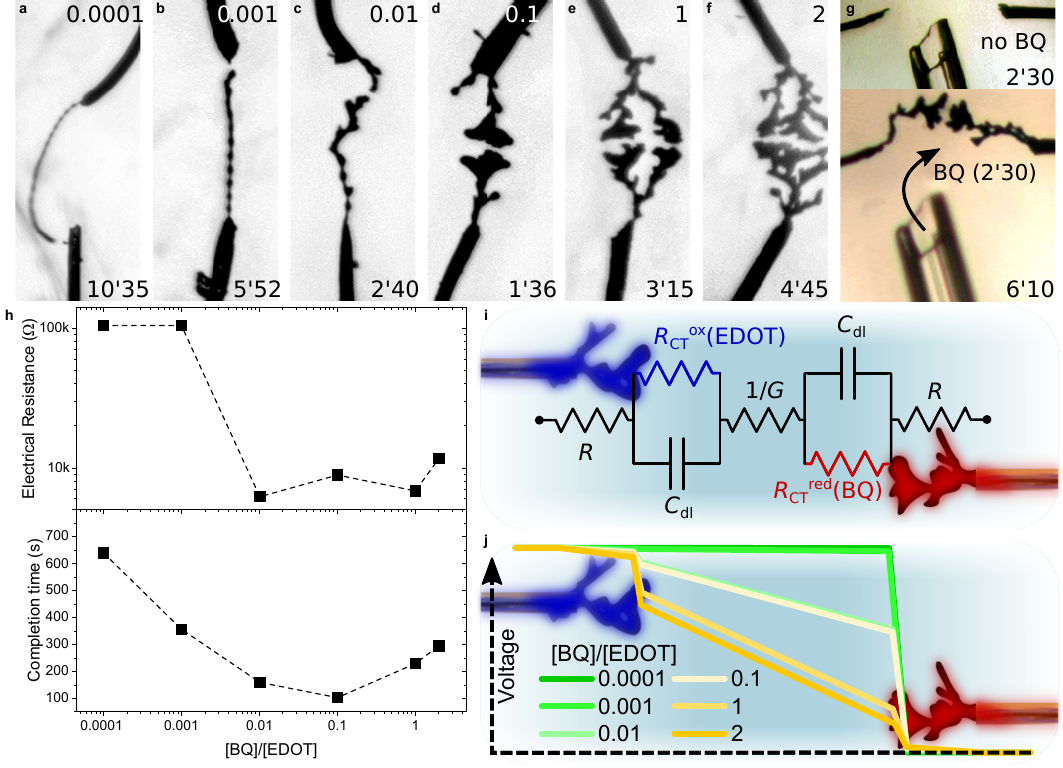}
  \caption{\textbf{Impact of BQ as an Electroactive Counter-Agent on the Electropolymerization of CPDs in Water $\vert$ a-f,} Optical pictures of CPD growths (4~V\textsubscript{p}, 0~V\textsubscript{off}, 80~Hz, 50\%\textsubscript{dc}, 10~mM [EDOT+BQ] and 1~mM NaPSS) in an aqueous media with different [BQ]/[EDOT] molar ratios (values displayed at the top of each picture). Each image was taken a few seconds before CPDs merged (applied-voltage duration displayed at the bottom of each picture). \textbf{g,} Introduction of a BQ aqueous solution, triggering CPD growth in an aqueous solution containing 10~mM EDOT and 1~mM NaPSS. \textbf{h,} Dependence of the electrical resistance (through each CPD displayed in Fig.\ref{fig:fig4}.a-f after they merged) and their growth duration on the [BQ]/[EDOT] molar ratio contained in the growth media. \textbf{i,} 7-Element equivalent circuit highlighting the contributions of [EDOT] and [BQ] on the charge-transfer balance between oxidation and reduction. \textbf{j,} Simulated voltage drop according to the equivalent circuit presented in Fig.\ref{fig:fig4}.i, with \textit{R} defined as half the experimental value of the CPD electrical resistance once merged, 1/\textit{G}~=~60~k$\Omega$, R\textsubscript{CT}\textsuperscript{ox}~=~100~k$\Omega$/mM\textsubscript{EDOT} and R\textsubscript{CT}\textsuperscript{red}~=~100~k$\Omega$/mM\textsubscript{BQ}.}
  \label{fig:fig4}
  \end{figure}

The electropolymerization of EDOT into PEDOT is an oxidation reaction that releases protons in the medium, thus requiring a counterbalancing reduction reaction to happen in the system to respect the electroneutrality of the environment. 
This aspect is often neglected, and the reduction is typically left to the solvent. \cite{Watanabe2024, Cucchi2021a} 
However, controlling growth calls for a complete understanding of both the anodic and the cathodic reaction. 
Even when water was used as the solvent, no gas production was evidenced, suggesting that Nernst thermodynamic potentials are not reliable predictors of Faradaic involvement of solvent molecules during CPD morphogenesis under an AC electric field --- including when the voltage amplitude exceeds the electrochemical window of the solvent. 
In that regard, Inagi and coworkers introduced p-benzoquinone (BQ) in acetonitrile to behave as a sacrificial reagent for the redox reaction and avoid pH changes due to the release of protons during electropolymerization \cite{Koizumi2016}. 
However, this behavior is not consistently observed when growing CPDs in acetonitrile (Table \ref{table:tab1}). 
Here, we propose to investigate the impact of the BQ:EDOT ratio on the growth and morphology of CPDs in water. 
To do so, the total amount of electroactive material in the medium was kept constant throughout this series of experiments ([BQ] + [EDOT] = 10~mM), so that the thermodynamics of the reaction would remain as similar as possible while the influence of BQ was under study.\\[3pt]
The influence of the BQ:EDOT ratio on CPD growth can be observed in Fig.\ref{fig:fig4}.a-f, where the impact on morphology and completion time can be appreciated. 
These results show that even small amounts of BQ (1:10000 and 1:1000 BQ:EDOT) enable dendritic formation. In this case, the electrogenerated fibers are thin and linear and take longer to form. 
Interestingly, these dendrites grew mainly out of one of the two electrodes of the system in a rosary-like fashion, rather than the simultaneous growth from both electrodes that can usually be observed. 
As the ratio increases up to 1:10, the completion time decreases and the objects progressively become more voluminous. 
For higher BQ contents, the dendrites keep getting more ramified. 
The increase in completion time observed for the 1:1 and 2:1 BQ:EDOT ratios might be due to the associated decrease in the monomer concentration rather than to the effect of ratio modulation itself, since the rather low solubility of EDOT in water prevented us from testing higher ratios under these experimental conditions. 
Indeed, as the BQ:EDOT ratio increases, the concentration of EDOT in the solution decreases as the quantity of electroactive species present in the environment was kept constant throughout this set of experiments.\\[3pt]
Understanding the influence of the electrochemical environment is precious asset to control growth, as illustrated by Fig.\ref{fig:fig4}.g, where the usual experimental setup as described above was used, except that BQ was removed from the system. 
In the absence of a sacrificial reagent, no growth was to be observed for over two minutes, although a signal was applied between the two electrodes. 
BQ was then gently introduced in the system with the help of a capillary, consequently triggering growth. 
This experiment suggests that CPD growth could be controlled by finely tuning the electrochemical medium by locally introducing small amounts of certain chemical species, a feature that could be critical to perform growth in specific environments such as biological ones, as many molecules (including BQ) are not biocompatible.\\[3pt]
The electrical measurements displayed in Fig.\ref{fig:fig4}.h once again highlight the dependence of the electrical resistance of the polymer fibers with their morphology: linear dendrites ([BQ]/[EDOT] $\leq$ 0.001) present a higher electrical resistance than more voluminous and ramified fibers ([BQ]/[EDOT] $\geq$ 0.001). 
A strong correlation can also be observed between the completion time of the CPDs and the BQ:EDOT ratio with a marked decrease as the ratio increases, although the increasing completion time for the last two points of the graph might be related to the low concentration of monomer in the solution.\\[3pt]
The equivalent circuit presented in Fig.\ref{fig:fig4}.i helps understanding how the BQ:EDOT ratio influences the electropolymerization process. 
During growth, a redox reaction happens at the electrode/electrolyte interfaces. 
In the equivalent circuit, this faradaic reaction is represented as a charge-transfer resistance. 
The influence of BQ and EDOT concentrations was assumed to be linear with respect to the charge-transfer resistance, such that R\textsubscript{CT}\textsuperscript{ox} = 100~k$\Omega$/mM\textsubscript{EDOT} and R\textsubscript{CT}\textsuperscript{red} = 100~k$\Omega$/mM\textsubscript{BQ}. 
Fig.\ref{fig:fig4}.j shows that for very low concentrations of BQ, the charge-transfer resistance at the cathode is extremely high, therefore causing most of the voltage to drop at the cathode/electrolyte interface. 
As [BQ]/[EDOT] increases, so does the value of R\textsubscript{CT}\textsuperscript{red}/R\textsubscript{CT}\textsuperscript{ox}, causing the voltage drops in the system to distribute more evenly and facilitating electropolymerization. 
BQ therefore has a similar impact on CPD growth than NaPSS (see Fig.\ref{fig:fig2}) as they both influence the drops of voltage at the interfaces, although they influence different parameters. 
However, the theoretical assumption of an increasing charge-transfer resistance for the reduction reaction with decreasing BQ concentration contradicts experimental observations, as growth was still observed at low BQ concentrations. 
A possible explanation is the presence of trace contaminants originating from the substrate, the electrolyte preparation, or the ambient environment. 
Experimentally, limiting the availability of BQ is a severe constrain for CPD growth, as the fibers grow slower and less bulky. 
These results confirm that introducing BQ as an oxidizing agent in trace amounts may be sufficient to trigger CPD growth. 
Even in water, the lack of control in the concentration of BQ as oxidizing counter-agent has an important influence over the growth kinetics and the resulting CPD morphologies.

\newpage

\subsection{Electrokinetic Control of the Growth Morphology by Viscosity Variation}

\begin{figure}
  \centering
  \includegraphics[width=1\columnwidth]{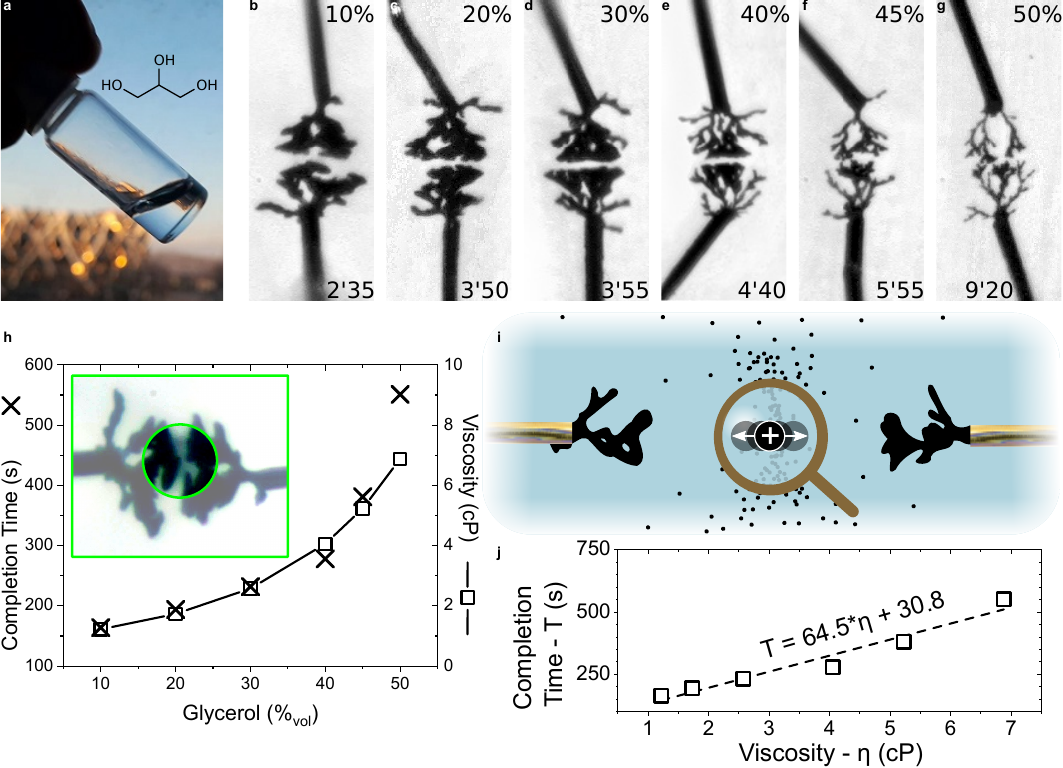}
  \caption{\textbf{Impact of Glycerol as a Higher-Viscosity Co-Solvent on the Development of CPDs in Water $\vert$ a,} Image of pure glycerol as a small-molecule non-acidic hydroxylated solvent with a higher viscosity than water (comparable to honey).\cite{Sheely1932,Yanniotis2006} \textbf{b-g,} Optical pictures of CPD growths (5~V\textsubscript{p}, 0~V\textsubscript{off}, 40~Hz, 50\%\textsubscript{dc}, 10~mM EDOT, 10~mM BQ and 1~mM NaPSS) in aqueous media with different volume contents of glycerol in water (values displayed at the top of each picture). Each image is taken a few seconds before CPD merge (applied-voltage duration displayed at the bottom of each picture). \textbf{h,} Dependence of the time taken to merge for each CPD displayed in Fig.\ref{fig:fig5}.b-g with the volume content of glycerol contained in the growth media (the trend is compared to viscosity values of glycerol/water mixtures calculated from Cheng \textit{et al.}).\cite{Cheng2008} As an inset, contrasted image of the 10\% growth where densities of flowing particles were evidenced in the video images. \textbf{i,} Schematic for a growth mechanism driven by the distribution of positively-charged conducting-polymer particles in a medium between electrodes polarized with a transient applied voltage. \textbf{j,} Correlation between the time taken to merge for each CPD displayed in Fig.\ref{fig:fig5}.b-g and the viscosity values displayed in \ref{fig:fig5}.h.}
  \label{fig:fig5}
  \end{figure}

Previously, we have seen the effect of selecting a different ion on the morphology of the CPD. 
By choosing emimOTf as a counter ion, the growth occurs extremely fast, and for the same growth frequency, the branches are thinner. 
It is hypothesized that this change of morphology is mostly due to differences in diffusion coefficients: triflate ions are light and easily migrate in water, while PSS ions are massive and will move slowly. 
One way to evaluate the changes of morphology with the diffusion coefficient without involving the chemistry of the ion is to change the viscosity of the environment, which can be achieved by introducing a co-solvent in water. 
As viscosity increases, electrokinetic effects are hindered, which could affect the morphology of the CPD, as it was observed that electrokinetic effects of possibly different natures (electrophoretic and electroconvective) appear at the voltage where CPDs start to grow, around 3.5~V, 80~Hz, in water. 
It has not yet been established whether CPDs can grow in a medium where electrokinetic effects are significantly reduced. 
Moreover, viscous media are often easier to integrate into complex systems due to their reduced volatility and more controlled flow. 
In water, viscosity can be modulated by adding a fraction of glycerol (Fig.\ref{fig:fig5}.a), a non toxic, non-acidic, hydroxylated small molecule with a viscosity more than a thousand times higher than water. 
It is also freely miscible with water, and mixtures of water/glycerol have been well studied in literature allowing viscosity values for different ratios to be estimated using existing models. 
Finally, it is expected that its addition does not alter the solvent's chemistry significantly. 
Electropolymerizations at different ratios of glycerol/water were conducted using the same concentration of monomer and ions than in the previous studies (1~mM NaPSS, 10~mM EDOT). 
Pictures of the resulting CPDs are presented Fig.\ref{fig:fig5}.b-g. A progression in the morphology is visible: for low ratios, branches will be thicker. 
Close to completion, they will adopt a unified face, where most of the branches will progress at the same time, very close to each other. 
In that situation, oligomer particles can be seen transiting between the two faces. 
A clear transition in morphology starts to appear around 40\%, where the thickness is notably reduced. 
Finally, the 50\% ratio gives thin branches that are more distant from each other. 
The time taken for the CPDs to connect to each other was recorded and plotted on Fig.\ref{fig:fig5}.h against the percentages of glycerol in solution. 
It shows a non linear increase in completion times which decently matches the increase of viscosity obtained for different mixtures of glycerol/water using the model of Cheng \textit{et al.}\cite{Cheng2008}, directly linking the dendrites' growth to viscosity. 
The change of viscosity between 40\% and 50\%, corresponding to the biggest changes in CPD morphology in the pictures, is also substantial compared to viscosities for percentages lower than 30\%. 
A previous work by Kumar \textit{et al.}\cite{Kumar2022} has shown that in the case of a low concentration electrolyte, charged particles subjected to an AC electric field between two rectangular electrodes tend to concentrate over time at the center of the gap between the electrodes. 
These charged particles can be oligomers whose electropolymerization reaction was interrupted by the periodic change of sign of the AC electric field, or parts of the already grown oxidized CPDs detached from the CPDs under the existing electric and hydrodynamic stresses. 
As shown in the inset of Fig.\ref{fig:fig5}.h, for thick CPDs close to completion, those particles are confined in the gap and visible as a black cloud constantly moving back and forth between the two sides, an indication that oligomer particles are present in solution and possibly reacting within the bulk. 
The closer the tips of the CPDs are to each other and the more likely these particles will be visible as the reaction reactants are all in close proximity, presumably allowing to reach higher degrees of polymerization; but it is also not excluded that those particles, for higher values of the gap between the tips, still exist but are not visible with an optical microscope. 
Their back-and-forth motion is depicted in Fig.\ref{fig:fig5}.i and is thought to be a form of AC-field driven electrophoresis\cite{Koizumi2016} of oligomers either dispersed or dissolved in solution. 
For particles confined in the gap, an increased viscosity means statistically less particles will reach the tips of the CPD for the same period of time, leading to thinner branches. 
The resulting morphology is similar to the one observed for high growth signal frequencies, except that in the case of adding glycerol as co-solvent, the growth is significantly slower when compared to bulky dendrites. 
Fig.\ref{fig:fig5}.j shows a clear correlation between the viscosity of the solution and the completion time of the CPDs, suggesting that the completion time could be predicted for a given viscosity based on the empiric linear relation displayed in the figure. 
However, it is important to stress that the completion time exhibits some level of variability as reported in water\cite{Janzakova2021a}. 
In conclusion, considering the limited window for electropolymerization offered by water where hydrolysis can happen for low frequencies below 20~Hz, viscosity is limiting the range of achievable morphologies to thin CPDs, but does not completely prevent growth in viscous media. 
These results stress the particular challenge for enabling CPD morphogenesis in solid or colloidal electrolytes such as hydrogels. 

\subsection{Control of Electroactive Monomers on the Structure of Conducting Polymer Dendrites}

\begin{figure}[!h]
  \centering
  \includegraphics[width=1\columnwidth]{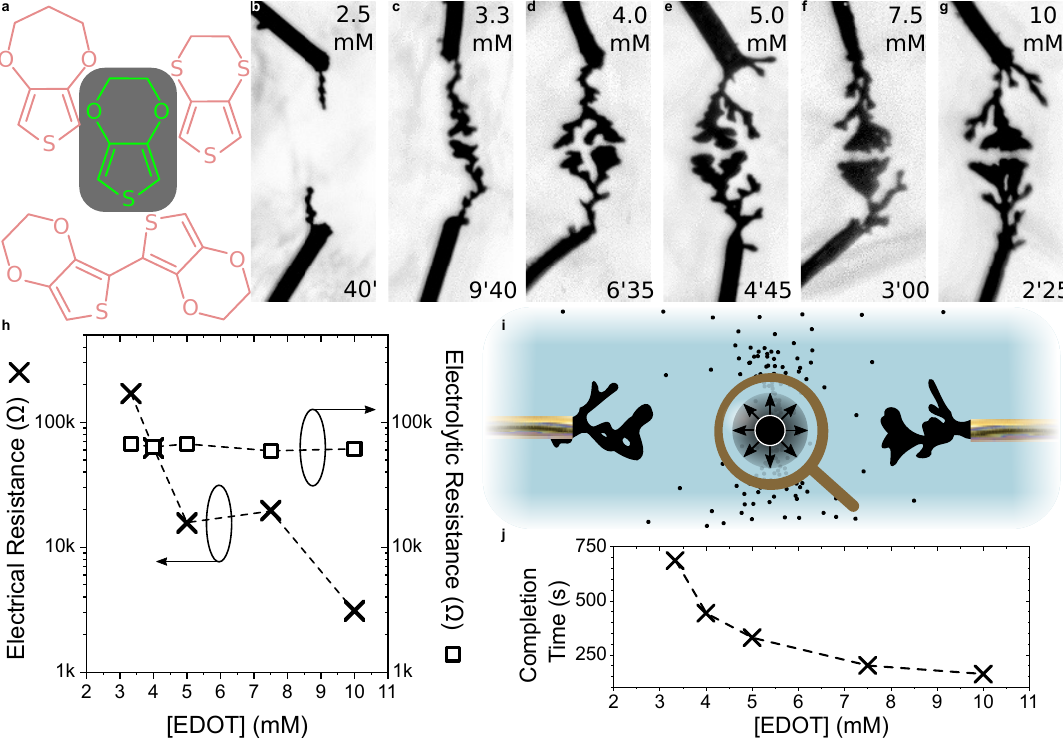}
  \caption{\textbf{Dependence of the Availability of EDOT as an Electroactive Monomer in an Aqueous Medium on CPD Morphogenesis $\vert$ a,} EDOT as a poorly but yet significantly more water soluble monomer than other structurally similar derivatives, such as 3,4-propylenedioxythiophene, 3,4-ethylenedithiathiophene or the dimer of EDOT, which are insoluble in water. \textbf{b-g,} Optical pictures of CPD growths (4~V\textsubscript{p}, 0~V\textsubscript{off}, 80~Hz, 50\%\textsubscript{dc}, 10~mM BQ and 1~mM NaPSS in water) with different molar concentrations of EDOT (values displayed at the top of each picture). Each image was taken a few seconds before CPDs merged (applied-voltage duration displayed at the bottom of each picture). \textbf{h,} Dependence of the electrical resistance (through each CPD displayed in Fig.\ref{fig:fig6}.b-g after they merged) and the electrolytic resistance (evaluated by impedancemetry at 10~kHz with a silver wire) on the concentration of EDOT in the growth media. \textbf{i,} Schematic of a growth mechanism limited by the concentration of EDOT where the kinetics of the reaction are influenced by the particle density and size distribution in the medium. \textbf{j,} Dependence of the completion time of each CPD displayed in Fig.\ref{fig:fig6}.b-g on the concentration of EDOT in the growth media.}
  \label{fig:fig6}
  \end{figure}

In electrochemistry, water is often disregarded as a solvent given its short electrochemical window. 
As a protic solvent, water induces pH-dependence in redox equilibria, necessitating the use of a buffer to maintain stability during the anodic electropolymerization of proton-releasing thiophene-based monomers. 
Additionally, many organic molecules are poorly soluble in water, which requires chemical engineering strategies to promote the hydrophilicity of thiophene-based monomers.\cite{Akoudad2000,Perepichka2002,Cutler2005,Stavrinidou2014,Mantione2020,Ghazal2023} In this regard, EDOT is an exceptional candidate: at the opposite of its dimer, its three-carbon dioxyalkylated analog, or its ethylenedithia equivalent (Fig.\ref{fig:fig6}.a), EDOT can be solubilized in pure water up to about 15~mM at 25°C.\cite{Schweiss2005} 
Watanabe \textit{et al.} demonstrated that a threefold more concentrated medium affects the morphology of electrogenerated fibers.\cite{Watanabe2018} 
However, the specific influence of monomer concentration on growth has never been characterized, despite the direct impact of EDOT availability on the formation of PEDOT-based CPDs. 
Given the low solubility of EDOT in water, even in the presence of surfactant, this parameter is therefore of particular importance. 
Investigating CPD morphogenesis in water with respect to monomer concentration is necessary to assess the feasibility of its implementation on water-based substrates.\\[3pt]

\begin{figure}
  \centering
  \includegraphics[width=1\columnwidth]{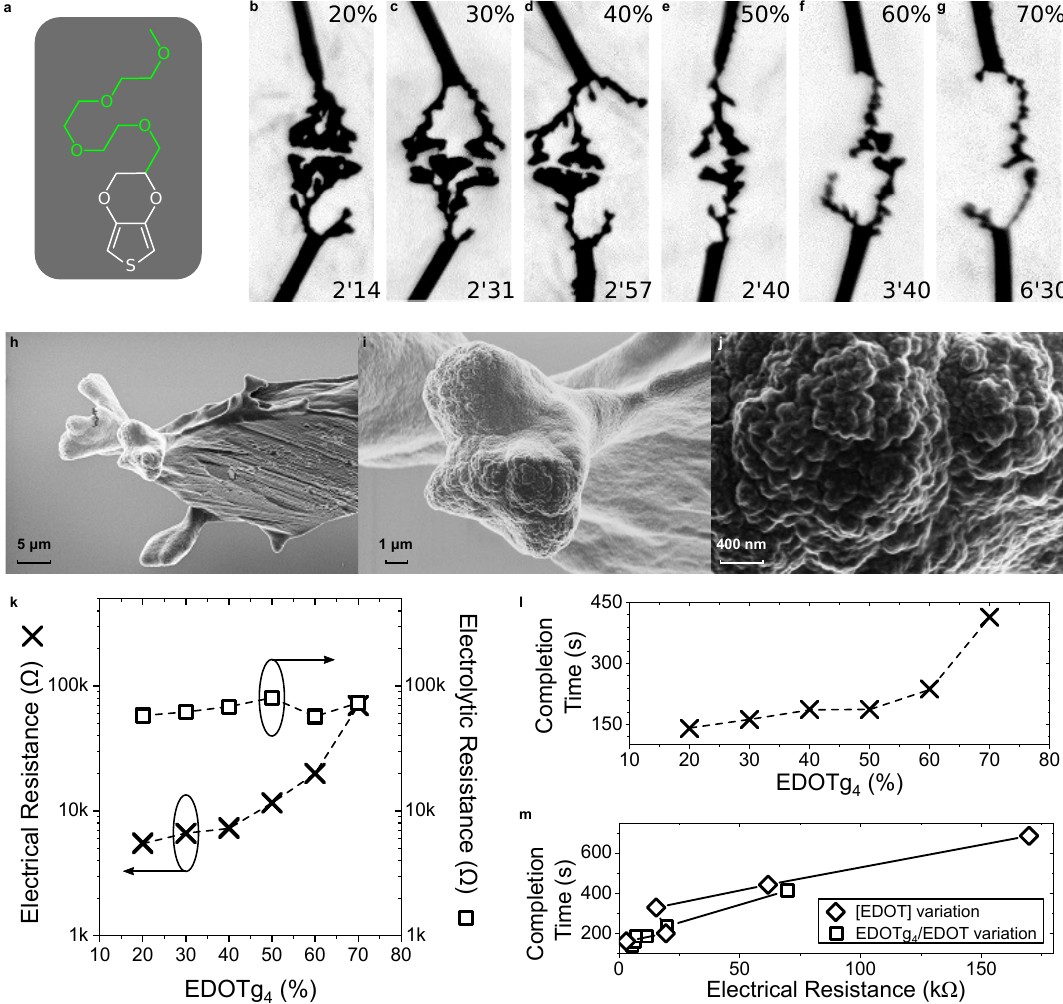}
  \caption{\textbf{Impact of an Increased Monomer Hydrophilicity with EDOTg\textsubscript{4} on the Development of CPDs in Water $\vert$ a,} Chemical structure of EDOTg\textsubscript{4}, as a structural analogue of EDOT with a hydrophilic side chain, synthesized as already reported by Ghazal \textit{et al}.\cite{Ghazal2023} \textbf{b-g,} Optical pictures of CPD growths (4~V\textsubscript{p}, 0~V\textsubscript{off}, 80~Hz, 50\%\textsubscript{dc}, 10~mM of total monomer content, 10~mM BQ and 1~mM NaPSS in water) with different molar ratios of EDOTg\textsubscript{4} copolymerized with EDOT (values displayed at the top of each picture). Each image was taken a few seconds before CPDs merged (applied-voltage duration displayed at the bottom of each picture). \textbf{h-j,} SEM images of a CPD growth in the same conditions as displayed in Fig.\ref{fig:fig7}.g. The morphology appears very similar to the one of a CPD grown in the same conditions but without EDOTg\textsubscript{4}. \textbf{k,} Dependence of the electrical resistance (through each CPD displayed in Fig.\ref{fig:fig7}.b-g after they merged) and the electrolytic resistance (evaluated by impedancemetry at 10~kHz with a silver wire) on the molar ratio of EDOTg\textsubscript{4} copolymerized with EDOT. \textbf{l,} Dependence of completion time of each CPD displayed in Fig.\ref{fig:fig7}.b-g on the molar ratio of EDOTg\textsubscript{4} copolymerized with EDOT. \textbf{m,} Comparison between the influence of EDOT dilution caused by EDOTg\textsubscript{4} substitution at 10~mM total content of monomers on growth latency and electrical resistance, and the results obtained in Fig.\ref{fig:fig6} for EDOT variation only. Both trends are comparable.}
  \label{fig:fig7}
  \end{figure}

Using the experimental setup described above, six dendrites were grown at different concentrations of EDOT (Fig.\ref{fig:fig6}). 
The influence of this concentration can be seen on the microscope pictures (Fig.\ref{fig:fig6}.b-g): as the quantity of monomer increases, the electrogenerated objects tend to become more ramified, although this effect is less pronounced than in the case of NaPSS (Fig.\ref{fig:fig2}) or BQ variations (Fig.\ref{fig:fig4}). 
Due to the low solubility of EDOT in water, only a narrow concentration range was attainable, which could partly explain the lack of noticeable morphological differences. 
The impact of monomer concentration on growth is best observed when looking at the completion time of the fibers. 
It appears easier for the dendrites to grow when more EDOT is present. 
When EDOT concentration was lower than 2.5~mM, the fibers could not merge after 40~min of growth, whereas the dendrites formed within just 145~s with 10~mM of EDOT. 
This parameter also had an impact on the conductance of the CPDs (Fig.\ref{fig:fig6}.h), as the electrical resistance of the fibers tended to decrease as monomer concentration increased. 
Once again, this observation could be explained by the morphology of the fibers rather than by an increased conductivity: since the electrolytic resistance appears to remain constant throughout the set of experiments (the concentration of NaPSS is set to 1~mM), it appears that the level of doping should be equivalent in all cases, and so should conductivity.\\[3pt]
That only a 4-time dilution can prevent growth suggests that the process is not linearly related to the concentration of monomer in the solution. 
PEDOT growth rate would be expected to be directly proportional to the available quantity of monomer in the solution. 
Yet, this is not what can be observed, as the completion time appears to exponentially increase as [EDOT] decreases (Fig.\ref{fig:fig6}.j). 
As the amount of BQ in the system is set to 10~mM, this parameter should not limit the reaction, as investigated in Fig.\ref{fig:fig4}. 
Interestingly, and assuming the concentrations of electrolyte  and BQ remain constant, lowering the concentration of EDOT should further increase the voltage drop at the anode interface, therefore promoting EDOT oxidation through an increased voltage bias in a two-electrode configuration. 
One hypothesis could be that the concentration of monomer conditions the density and size distribution of the particles in the environment, thus electrokinetically limiting the reaction (Fig.\ref{fig:fig6}.i). 
The relationship between EDOT concentration and the colloidal properties of the medium as a suspension of electrogenerated PEDOT particules is not obvious and would require simulations. 
However, it appears at the moment to be the most plausible reason to explain the superlinearity of CPD growth with EDOT concentration by the closeness of PEDOT particles to condition their probability to coalescence in a volume of medium in their free state, indirectly impacting the CPD growth on the gold wires.\\[3pt]
The challenge imposed by the low solubility of thiophene-based monomers in water may be addressed by chemically engineering the molecules. 
As shown by Koizumi \textit{et al.},\cite{Koizumi2016} morphology can be controlled by modifying the chemistry on an EDOT monomer substituted on the ethylenedioxy bridge, allowing to change its environment affinity without altering the electroactivity of the monomer. 
In the same conditions, substituting EDOT on the ethylenedioxy bridge promotes less dendritic and longer-branched growths with a -C\textsubscript{10}H\textsubscript{21} alkylchain substitution than with a -CH\textsubscript{3} substitution.\cite{Koizumi2016} 
Yet, this effect might not be observed with any substitute of EDOT. 
A derivative of EDOT with a glycolated side chain, EDOTg\textsubscript{4},\cite{Ghazal2023} was introduced to be co-electropolymerized alongside EDOT (Fig.\ref{fig:fig7}.a). 
The total amount of monomer in the solution was kept constant ([EDOT] + [EDOTg\textsubscript{4}] = 10~mM), while the proportion of EDOTg\textsubscript{4} in the solution was growing. 
Fig.\ref{fig:fig7}.b-g show optical pictures of polymer fibers grown in the presence of both EDOT and EDOTg\textsubscript{4}, from 20\% EDOTg\textsubscript{4} to 70\% EDOTg\textsubscript{4}. 
As it can be observed, the electrogenerated fibers tended to become more linear as the proportion of EDOTg\textsubscript{4} in the solution increased, and the completion time of the dendrites rose. 
Contrary to the experiments performed by Koizumi, Watanabe \textit{et al.} on the effect of EDOT substitution,\cite{Koizumi2016,Watanabe2018} no growth was observed in the case EDOTg\textsubscript{4} was used as the only monomer. 
This correlates with Ghazal \textit{et al.}'s results showing that DC electro-co-polymerization of EDOT with EDOTg\textsubscript{4} above 70\% of EDOTg\textsubscript{4} produces a polymer that is too hydrophilic to remain on the electrodes.\cite{Ghazal2023} 
As a result, it is difficult in the current state to attribute the change in morphology to either the presence of EDOTg\textsubscript{4} or the dilution of EDOT, as it has previously been evidenced to impact growth morphology. 
The fibers did not present any visual clue of the presence of EDOTg\textsubscript{4} within the polymerized structure. 
The SEM images displayed Fig.\ref{fig:fig7}.h-j also confirm that the morphology of these fibers does not strongly differ from that of dendrites grown only in the presence of EDOT (see Fig.\ref{fig:fig3}.c-e). 
In addition, the electrical resistance of these structures tended to increase with the amount of EDOTg\textsubscript{4} in the medium, whereas the electrolytic resistance remained constant, thus discarding the possibility of a decrease in conductivity (Fig.\ref{fig:fig7}.k). 
The completion time of these dendrites also increased importantly with the [EDOTg\textsubscript{4}]/[EDOT] ratio (Fig.\ref{fig:fig7}.l). 
From these observations, it appears that the presence of EDOTg\textsubscript{4} within the electropolymerized dendrites cannot be confirmed. 
Moreover, the results discussed above are highly similar to those obtained when only decreasing the concentration of EDOT, notably the electrical resistance and completion time of the fibers, as can be observed in Fig.\ref{fig:fig7}.m. 
It therefore appears that EDOTg\textsubscript{4} could act only as a spectator to the reaction, showing that modifying the chemistry of the monomer might not be sufficient to influence dendritic development. 

\begin{figure}
  \centering
  \includegraphics[width=1\columnwidth]{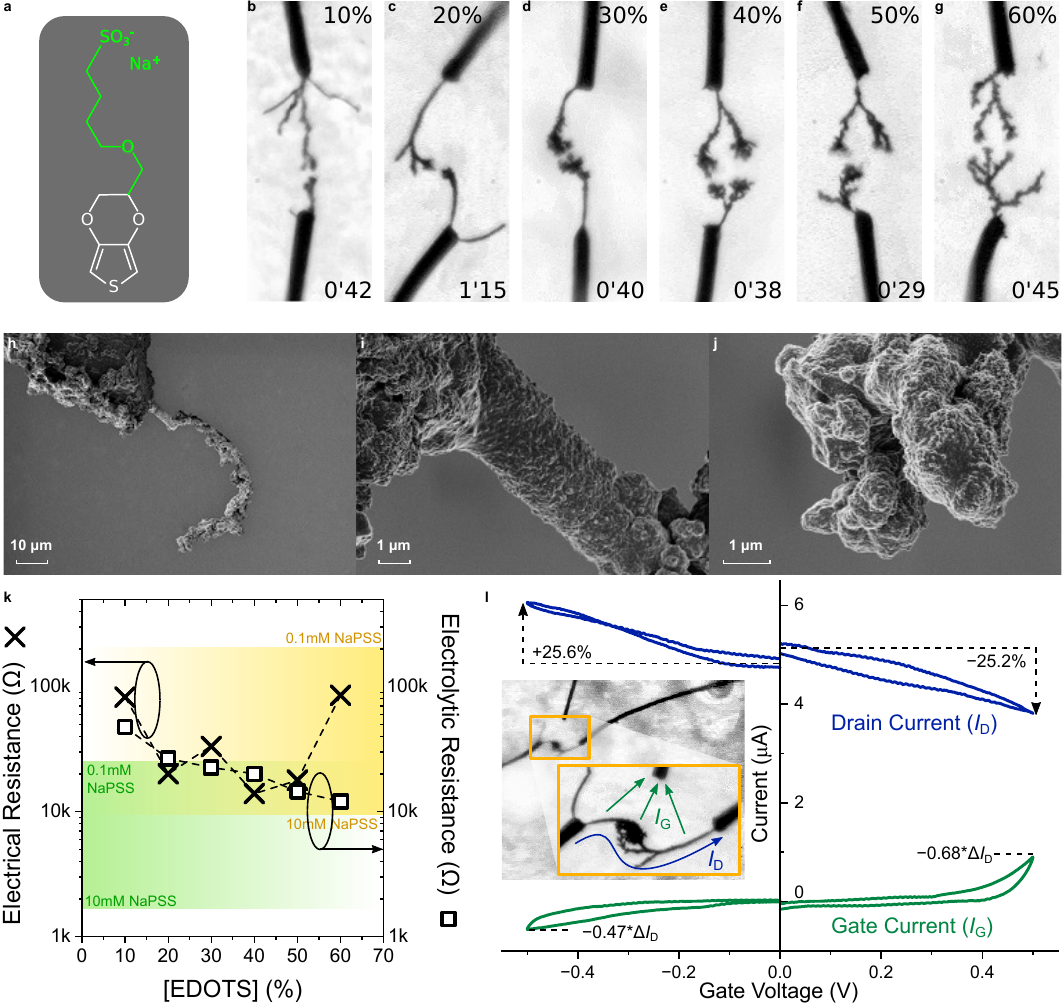}
  \caption{\textbf{Impact of EDOT-S as an Electroactive Conjugated Polyelectrolyte Precursor on the Development of CPDs in Water $\vert$ a,} Chemical structure of EDOT-S, as a structural analogue of EDOT and as an electrolyte. \textbf{b-g,} Optical pictures of CPD growths (4~V\textsubscript{p}, 0~V\textsubscript{off}, 80~Hz, 50\%\textsubscript{dc}, 10~mM of total monomer content and 10~mM BQ in water) with different molar ratios of EDOT-S copolymerized with EDOT (values displayed at the top of each picture). Each image is taken a few seconds before CPDs merge (applied-voltage duration displayed at the bottom of each picture). \textbf{h-j,} SEM images of a CPD growth in the same conditions as displayed in Fig.\ref{fig:fig8}.f. The morphology exhibits a highly distinct structure from the one of a CPD grown in the same conditions but without EDOT-S content. \textbf{k,} Dependence of the electrical resistance (through each CPD displayed in Fig.\ref{fig:fig8}.b-g after they merge) and the electrolytic resistance (evaluated by impedancemetry at 10~kHz with a silver wire) on the molar ratio of EDOT-S copolymerized with EDOT. \textbf{l,} Transfer characteristic of the CPDs displayed in Fig.\ref{fig:fig8}.b-g but in a merged state, using a silver wire as an electrochemical gate electrode (as displayed in the inset). The transistor characterization shows a transistor effect both in accumulation and depletion mode, attesting from a mild doping of the CPDs due to the presence of EDOT-S monomer units in the material.}
  \label{fig:fig8}
  \end{figure}

In addition to the solvent, the CPD growths conducted so far always involved three ingredients: a monomer, an electrolyte and a redox agent. 
Preparing complex media from multiple constituents may simultaneously induce different variabilities that can all impact the growth. 
To address these challenges, efforts have been made to simplify the composition of the medium with components assuming multiple roles at the same time.\\[3pt]
First, the electrolyte can act as a solvent itself, like in the case of ionic liquids which showed successful dendritic growth.\cite{Chen2023} When the salt is also an oxidizing agent, the electrolyte can even substitute BQ as a redox agent supporting EDOT to PEDOT oxidation, such as H\textsubscript{2}PtCl\textsubscript{6}\cite{Koizumi2018}. 
Here, the use of a conjugated polyelectrolyte precursor as both a monomer and an electrolyte is investigated. 
Sodium 4-(2,3-dihydrothieno[3,4-b][1,4]-
dioxin-2-yl-methoxy)-1-butanesulfonate (EDOT(RSO\textsubscript{3}Na) or EDOT-S is an electroactive EDOT monomer derivative, carrying a sulfonate-terminated alkoxy chain on the ethylene covalently attached dioxy bridge (Fig.\ref{fig:fig8}.a).\cite{Stephan1998,Cutler2005,Yano2019} 
This leads to a greater solubility for the monomer in water compared to regular EDOT with a sulfonate group, carrying a negative charge to support the polymer p-doping. 
Because of the sulfonate-terminated charge on the substituted monomer, the effect of the substitution may be significantly different compared to the impact of a neutral glycol chain on EDOTg\textsubscript{4}.\\[3pt] 
Similarly to EDOTg\textsubscript{4}, EDOT-S was electro-co-polymerized with EDOT monomers using different ratios of EDOT-S at 10~mM of total monomer, same as for benzoquinone. No extra salts were involved; thus, the electrolyte is composed of EDOT-S at different concentrations, supported by sodium cations. 
The morphology of a CPD depends on the rate of EDOT-S involved during electro-co-polymerization (Fig.\ref{fig:fig8}.b-g). 
At 10\% (10~mM monomer with 1~mM EDOT-S in Fig.\ref{fig:fig8}.b), the branches are thinner and more linear compared to 10~mM EDOT growing in 1~mM NaPSS (Fig.\ref{fig:fig2}.f). 
This demonstrates that the anchoring of a sulfonate on monomers provides very different features to the morphogenesis than in the case of sulfonate-free monomers with PSS at the same concentration levels. 
At any EDOT-S:EDOT ratio, CPDs grow significantly faster. 
The fast, linear growth and thin thickness of branches bring them closer to PEDOT:OTf CPDs growth (Fig.\ref{fig:fig3}.b). 
However, a notable difference is the absence of branches on the sides of the wires. 
Increasing the fraction of EDOT-S between 20 and 40\% shows smooth curved stems, possibly influenced by electrohydrodynamic vortices arising under a highly confined geometry and during high voltage electrodeposition, that will favor circular trajectories. 
A multitude of small branches starts to appear at the tip of the CPDs as they get closer to each other. 
At ratios 50\% and 60\%, the CPD shows larger branches with an increased roughness, covering the entire CPD. 
At 60\% and above, the branches appear more transparent. 
Also, the initial coating enveloping the wire before the CPDs growth is seen progressing along the sides of the wire, reaching the tip before finally branching out as a CPD. This behaviour is usually not observed: ordinarily, either there is no coating on the wire or it appears uniformly at the beginning of the electropolymerization. 
Eventually, at higher ratios, bubbles will start to form and no growth is observed (the wires' color and thickness remain unchanged), unless wires are initially placed closer to each other. 
Attempting to electrochemically form an EDOT-S homopolymer is notoriously difficult in water, as it was first observed by St\'{e}phan \textit{et al.},\cite{Stephan1998} where its electropolymerization only resulted in soluble oligomers diffusing away from the electrodes.\\[3pt]
SEM images of EDOT-S CPDs grown for a ratio of 50\% of each monomer are shown in Fig.\ref{fig:fig8}.h-j. 
The coating on the wire presents a high roughness, that is also visible on the curved stem of the CPD. 
The start of the CPD, at the tip of the wire, is notably smooth and linear, reminiscent of the previously seen PEDOT:PSS CPDs' external surface. 
An abrupt change of morphology occurs shortly after and the CPD starts curving. 
The InLens signal gives differences of brightness between parts of the CPD that could potentially signify local differences in conductivity.\\[3pt]
The electrical and electrolytic resistances of the CPD have been measured after the CPDs were connected to each other, for the different proportions of EDOT and EDOT-S (Fig.\ref{fig:fig8}.k). 
As expected, the electrolytic conductance is controlled by the concentration of EDOT-S present in the medium. 
In contrast, the electric resistance does not vary monotonically with the ratio of EDOT-S:EDOT. 
There seems to be an optimum between lower concentrations of EDOT-S where a lower content of sulfonate groups promotes lower CPD conductances, and higher concentrations of EDOT-S for which the sulfonation of the CPDs may influence their fragility in water as charged particulate matter in a polar solvent. 
The lowest resistance is achieved for 40\% EDOT-S, which is still an order of magnitude higher than EDOT-based CPDs grown with 1~mM NaPSS in the same conditions. 
As an OMIEC material,\cite{Wan2022,Danielsen2022,Chae2024} a balance between electric and ionic conduction may control the electrochemical behavior of CPDs formed using EDOT-S copolymerized with EDOT. 
PEDOT CPDs act as organic electrochemical transistors (OECT) in aqueous electrolytes.\cite{Janzakova2021,Cucchi2021a,Scholaert2022} A similar behavior has also been evidenced in the case of CPDs copolymerized using EDOT and EDOT-S with no involvement of any other salts (Fig.\ref{fig:fig8}.l). 
As a conjugated polyelectrolyte, PEDOT-S has an inherent OECT effect.
\cite{Zeglio2015,Zeglio2016,Zeglio2017,NguyenDang2022,Llanes2023}

This effect was tested using a three-electrode setup consisting of two gold electrodes short-circuited by a EDOT-S/EDOT CPD, and a silver wire serving as a gate electrode, positioned close to the junction between the two CPDs. 
The transfer characteristic for a ratio of 20\% is shown Fig.\ref{fig:fig8}.l. Other ratio values were tested but yielded no significant differences. 
At this composition, the electrolytic resistance is only slightly greater than the electrical resistance of the channel. 
In this setup, a few factors work against the observation of an OECT effect: first, the electrolyte is in very low concentration (10~mM), secondly, the gate is uncoated and of identical diameter compared to the gold wires: the coated wires and CPD have higher surface area, the limited area of the gate therefore prevent additional ions from entering the bulk of the CPD. 
Despite these limitations, a transistor-like modulation is observed in both apparent accumulation (\textit{V}\textsubscript{G} < 0~V) and depletion (\textit{V}\textsubscript{G} > 0~V) modes. 
Visible signs of faradaic activity in the gate current only appear outside the [{\textminus}100 mV, +300 mV] potential window and remain lower than the drain current modulation, indicating a transistor-like effect of EDOT-S-based CPDs, achieved without the use of a supporting electrolyte in the aqueous growth medium.

\newpage

\section{Conclusion}
The ability for CPDs to grow in water has been studied by varying quantitatively the major components required for their growth. 
It was observed that they require the presence of ions from tenths of micromolars to tenths of millimolars, and that their chemical nature greatly impacts the growth kinetics and morphology of the CPDs. 
Even in only a minor amount compared to the monomer, a redox agent is always required to control growth: introducing this agent can trigger the growth mechanism. 
Even the addition of a cosolvent may affect growth in water. 
Monomers shall be in a high enough concentration to observe growth: at least 3~mM. As the solubility of EDOT is limited in water, hydrophilic EDOT derivatives may be involved in the growth process. 
While the effect of a glycol chain on EDOT does not promote significant changes in growth morphology and kinetics, a sulfonated chain promotes self-doping of the CPD, enabling genuine transistor effect without having to involve a source of ions other than the monomer salt itself. 
The exploration performed in this study concludes that implementing CPD morphogenesis in water is highly versatile and adaptable to very different aqueous environments, mimicking a natural adaptive mechanism for growing transistors in a drop of water, with minimalist fabrication resources and outside of a cleanroom environment.

\section*{Acknowledgments}
The authors thank the French National Nanofabrication Network \href{https://www.renatech.org/en/}{RENATECH} for financial support of the IEMN cleanroom. 
We thank also the IEMN cleanroom staff for their advices and support. 
This work is funded by ANR-JCJC "Sensation" project (grant number: \href{https://anr.fr/Projet-ANR-22-CE24-0001}{ANR-22-CE24-0001}), the ERC-CoG "IONOS" project (grant number: \href{https://doi.org/10.3030/773228}{773228}) and the R{\'e}gion Hauts-de-France.

\section*{Competing Interests}
The authors declare no competing interests.

\bibliographystyle{natsty-doilk-on-jour}  
\bibliography{ref}  

\begin{thebibliography}{10}
\expandafter\ifx\csname url\endcsname\relax
  \def\url#1{\texttt{#1}}\fi
\expandafter\ifx\csname urlprefix\endcsname\relax\def\urlprefix{URL }\fi
\providecommand{\bibinfo}[2]{#2}
\providecommand{\eprint}[2][]{\url{#2}}

\bibitem{Cisco2017}
\bibinfo{title}{{The Zettabyte Era: Trends and Analysis}}.
\newblock \bibinfo{type}{Tech. Rep.}, \bibinfo{institution}{Cisco}
  (\bibinfo{year}{2017}).
\newblock
  \urlprefix\url{https://www.hit.bme.hu/~jakab/edu/HTI18/Litr/Cisco_The_Zettabyte_Era_2017June__vni-hyperconnectivity-wp.pdf}.

\bibitem{Reinsel2017}
\bibinfo{author}{Reinsel, D.}, \bibinfo{author}{Gantz, J.} \&
  \bibinfo{author}{Rydnin, J.}
\newblock \bibinfo{title}{Data age 2025}.
\newblock \bibinfo{type}{Tech. Rep.} (\bibinfo{year}{2017}).
\newblock
  \urlprefix\url{https://www.seagate.com/files/www-content/our-story/trends/files/Seagate-WP-DataAge2025-March-2017.pdf}.

\bibitem{deVries2023}
\bibinfo{author}{{de Vries}, A.}
\newblock \bibinfo{title}{The growing energy footprint of artificial
  intelligence}.
\newblock
  \href{http://dx.doi.org/10.1016/j.joule.2023.09.004}{\emph{\bibinfo{journal}{Joule}}}
  \textbf{\bibinfo{volume}{7}}, \bibinfo{pages}{2191–2194}
  (\bibinfo{year}{2023}).

\bibitem{EPRI2024}
\bibinfo{title}{Powering intelligence - analyzing artificial intelligence and
  data center energy consumption}.
\newblock \bibinfo{type}{Tech. Rep.}, \bibinfo{institution}{Electric Power
  Research Institute (EPRI)} (\bibinfo{year}{2024}).
\newblock \urlprefix\url{https://www.epri.com/research/products/3002028905}.

\bibitem{IEA2025}
\bibinfo{title}{Energy and {AI}}.
\newblock \bibinfo{type}{Tech. Rep.}, \bibinfo{institution}{International
  Energy Agency (IEA)}, \bibinfo{address}{Paris} (\bibinfo{year}{2025}).
\newblock \urlprefix\url{www.iea.org/reports/energy-and-ai}.

\bibitem{Chen2025}
\bibinfo{author}{Chen, S.}
\newblock \bibinfo{title}{Data centres will use twice as much energy by 2030
  — driven by {AI}}.
\newblock
  \href{http://dx.doi.org/10.1038/d41586-025-01113-z}{\emph{\bibinfo{journal}{Nature}}}
   (\bibinfo{year}{2025}).

\bibitem{Pollard2021}
\bibinfo{author}{Pollard, J.}, \bibinfo{author}{Osmani, M.},
  \bibinfo{author}{Cole, C.}, \bibinfo{author}{Grubnic, S.} \&
  \bibinfo{author}{Colwill, J.}
\newblock \bibinfo{title}{A circular economy business model innovation process
  for the electrical and electronic equipment sector}.
\newblock
  \href{http://dx.doi.org/10.1016/j.jclepro.2021.127211}{\emph{\bibinfo{journal}{J.
  Clean. Prod.}}} \textbf{\bibinfo{volume}{305}}, \bibinfo{pages}{127211}
  (\bibinfo{year}{2021}).

\bibitem{Solomon2000}
\bibinfo{author}{Solomon, R.}, \bibinfo{author}{Sandborn, P.} \&
  \bibinfo{author}{Pecht, M.}
\newblock \bibinfo{title}{Electronic part life cycle concepts and obsolescence
  forecasting}.
\newblock
  \href{http://dx.doi.org/10.1109/6144.888857}{\emph{\bibinfo{journal}{IEEE
  Transactions on Components and Packaging Technologies}}}
  \textbf{\bibinfo{volume}{23}}, \bibinfo{pages}{707–717}
  (\bibinfo{year}{2000}).

\bibitem{Huang2019}
\bibinfo{author}{Huang, C.-M.}, \bibinfo{author}{Romero, J.~A.},
  \bibinfo{author}{Osterman, M.}, \bibinfo{author}{Das, D.} \&
  \bibinfo{author}{Pecht, M.}
\newblock \bibinfo{title}{Life cycle trends of electronic materials, processes
  and components}.
\newblock
  \href{http://dx.doi.org/10.1016/j.microrel.2019.05.023}{\emph{\bibinfo{journal}{Microelectron.
  Reliab.}}} \textbf{\bibinfo{volume}{99}}, \bibinfo{pages}{262–276}
  (\bibinfo{year}{2019}).

\bibitem{Nunes2021}
\bibinfo{author}{Nunes, I.~C.}, \bibinfo{author}{Kohlbeck, E.},
  \bibinfo{author}{Beuren, F.~H.}, \bibinfo{author}{Fagundes, A.~B.} \&
  \bibinfo{author}{Pereira, D.}
\newblock \bibinfo{title}{Life cycle analysis of electronic products for a
  product-service system}.
\newblock
  \href{http://dx.doi.org/10.1016/j.jclepro.2021.127926}{\emph{\bibinfo{journal}{J.
  Clean. Prod.}}} \textbf{\bibinfo{volume}{314}}, \bibinfo{pages}{127926}
  (\bibinfo{year}{2021}).

\bibitem{Vasan2014}
\bibinfo{author}{Vasan, A.}, \bibinfo{author}{Sood, B.} \&
  \bibinfo{author}{Pecht, M.}
\newblock \bibinfo{title}{Carbon footprinting of electronic products}.
\newblock
  \href{http://dx.doi.org/10.1016/j.apenergy.2014.09.074}{\emph{\bibinfo{journal}{Appl.
  Energy}}} \textbf{\bibinfo{volume}{136}}, \bibinfo{pages}{636–648}
  (\bibinfo{year}{2014}).

\bibitem{Crawford2019}
\bibinfo{author}{Crawford, K.} \& \bibinfo{author}{Joler, V.}
\newblock \bibinfo{title}{Anatomy of an ai system} (\bibinfo{year}{2019}).
\newblock
  \urlprefix\url{https://collections.vam.ac.uk/item/O1500029/anatomy-of-an-ai-system-poster-kate-crawford/}.

\bibitem{Dhar2020}
\bibinfo{author}{Dhar, P.}
\newblock \bibinfo{title}{The carbon impact of artificial intelligence}.
\newblock
  \href{http://dx.doi.org/10.1038/s42256-020-0219-9}{\emph{\bibinfo{journal}{Nat.
  Mach. Intell.}}} \textbf{\bibinfo{volume}{2}}, \bibinfo{pages}{423–425}
  (\bibinfo{year}{2020}).

\bibitem{Jones2014}
\bibinfo{author}{Jones, G.~G.}, \bibinfo{author}{Hanson, L.~M.} \&
  \bibinfo{author}{Kleist, T.}
\newblock \bibinfo{title}{Mobile communication device having multiple,
  interchangeable second devices} (\bibinfo{year}{2014}).
\newblock \urlprefix\url{https://patentcenter.uspto.gov/applications/12726252}.

\bibitem{Schischke2016}
\bibinfo{author}{Schischke, K.}, \bibinfo{author}{Proske, M.},
  \bibinfo{author}{Nissen, N.~F.} \& \bibinfo{author}{Lang, K.-D.}
\newblock \bibinfo{title}{Modular products: Smartphone design from a circular
  economy perspective}.
\newblock In \emph{\bibinfo{booktitle}{2016 Electronics Goes Green 2016+
  (EGG)}}, \bibinfo{pages}{1–8} (\bibinfo{year}{2016}).

\bibitem{Hankammer2018}
\bibinfo{author}{Hankammer, S.}, \bibinfo{author}{Jiang, R.},
  \bibinfo{author}{Kleer, R.} \& \bibinfo{author}{Schymanietz, M.}
\newblock \bibinfo{title}{Are modular and customizable smartphones the future,
  or doomed to fail? a case study on the introduction of sustainable consumer
  electronics}.
\newblock
  \href{http://dx.doi.org/10.1016/j.cirpj.2017.11.001}{\emph{\bibinfo{journal}{CIRP
  J. Manuf. Sci. Technol.}}} \textbf{\bibinfo{volume}{23}},
  \bibinfo{pages}{146–155} (\bibinfo{year}{2018}).

\bibitem{Salahuddin2018}
\bibinfo{author}{Salahuddin, S.}, \bibinfo{author}{Ni, K.} \&
  \bibinfo{author}{Datta, S.}
\newblock \bibinfo{title}{The era of hyper-scaling in electronics}.
\newblock
  \href{http://dx.doi.org/10.1038/s41928-018-0117-x}{\emph{\bibinfo{journal}{Nat.
  Electron.}}} \textbf{\bibinfo{volume}{1}}, \bibinfo{pages}{442–450}
  (\bibinfo{year}{2018}).

\bibitem{Zebulum2001}
\bibinfo{author}{Zebulum, R.~S.}, \bibinfo{author}{Pacheco, M.~A.} \&
  \bibinfo{author}{Vellasco, M. M.~B.}
\newblock \emph{\bibinfo{title}{{Evolutionary Electronics (1st edition)}}}
  (\bibinfo{publisher}{CRC Press}, \bibinfo{year}{2001}).

\bibitem{Yao1999}
\bibinfo{author}{Yao, X.} \& \bibinfo{author}{Higuchi, T.}
\newblock \bibinfo{title}{Promises and challenges of evolvable hardware}.
\newblock
  \href{http://dx.doi.org/10.1109/5326.740672}{\emph{\bibinfo{journal}{IEEE
  Trans. Syst. Man. Cybern. C Appl. Rev.}}} \textbf{\bibinfo{volume}{29}},
  \bibinfo{pages}{87–97} (\bibinfo{year}{1999}).

\bibitem{Haddow2011}
\bibinfo{author}{Haddow, P.} \& \bibinfo{author}{Tyrrell, A.}
\newblock \bibinfo{title}{Challenges of evolvable hardware: past, present and
  the path to a promising future}.
\newblock
  \href{http://dx.doi.org/10.1007/s10710-011-9141-6}{\emph{\bibinfo{journal}{Genet.
  Program. Evolvable Mach.}}} \textbf{\bibinfo{volume}{12}},
  \bibinfo{pages}{183–215} (\bibinfo{year}{2011}).

\bibitem{Stepanyants2002}
\bibinfo{author}{Stepanyants, A.}, \bibinfo{author}{Hof, P.~R.} \&
  \bibinfo{author}{Chklovskii, D.~B.}
\newblock \bibinfo{title}{Geometry and structural plasticity of synaptic
  connectivity}.
\newblock
  \href{http://dx.doi.org/10.1016/S0896-6273(02)00652-9}{\emph{\bibinfo{journal}{Neuron}}}
  \textbf{\bibinfo{volume}{34}}, \bibinfo{pages}{275–288}
  (\bibinfo{year}{2002}).

\bibitem{Chaiwanon2016}
\bibinfo{author}{Chaiwanon, J.}, \bibinfo{author}{Wang, W.},
  \bibinfo{author}{Zhu, J.-Y.}, \bibinfo{author}{Oh, E.} \&
  \bibinfo{author}{Wang, Z.-Y.}
\newblock \bibinfo{title}{Information integration and communication in plant
  growth regulation}.
\newblock
  \href{http://dx.doi.org/10.1016/j.cell.2016.01.044}{\emph{\bibinfo{journal}{Cell}}}
  \textbf{\bibinfo{volume}{164}}, \bibinfo{pages}{1257–1268}
  (\bibinfo{year}{2016}).

\bibitem{Tero2010}
\bibinfo{author}{Tero, A.} \emph{et~al.}
\newblock \bibinfo{title}{Rules for biologically inspired adaptive network
  design}.
\newblock
  \href{http://dx.doi.org/10.1126/science.1177894}{\emph{\bibinfo{journal}{Science}}}
  \textbf{\bibinfo{volume}{327}}, \bibinfo{pages}{439–442}
  (\bibinfo{year}{2010}).

\bibitem{Adamatzky2010}
\bibinfo{author}{Adamatzky, A.}
\newblock \emph{\bibinfo{title}{Physarum Machines}} (\bibinfo{publisher}{World
  Scientific}, \bibinfo{year}{2010}).

\bibitem{Adamatzky2019}
\bibinfo{author}{Adamatzky, A.}
\newblock \bibinfo{title}{A brief history of liquid computers}.
\newblock
  \href{http://dx.doi.org/10.1098/rstb.2018.0372}{\emph{\bibinfo{journal}{Philos.
  Trans. R. Soc. Lond. B Biol. Sci.}}} \textbf{\bibinfo{volume}{374}},
  \bibinfo{pages}{20180372} (\bibinfo{year}{2019}).

\bibitem{Maass2001}
\bibinfo{author}{Maass, W.}
\newblock \bibinfo{title}{wetware ({English} version)}.
\newblock In \emph{\bibinfo{booktitle}{{TAKEOVER}: {Who} is {Doing} the {Art}
  of {Tomorrow} ({Ars} {Electronica} 2001)}}, \bibinfo{pages}{148--152}
  (\bibinfo{publisher}{Springer}, \bibinfo{year}{2001}).
\newblock \urlprefix\url{/maass/129/129a.htm}.

\bibitem{Jung2023}
\bibinfo{author}{Jung, W.-B.} \emph{et~al.}
\newblock \bibinfo{title}{An aqueous analog mac machine}.
\newblock
  \href{http://dx.doi.org/10.1002/adma.202205096}{\emph{\bibinfo{journal}{Adv.
  Mater.}}} \textbf{\bibinfo{volume}{35}}, \bibinfo{pages}{2205096}
  (\bibinfo{year}{2023}).

\bibitem{Sato2023}
\bibinfo{author}{Sato, D.} \emph{et~al.}
\newblock \bibinfo{title}{Characterization of information-transmitting
  materials produced in ionic liquid-based neuromorphic electrochemical devices
  for physical reservoir computing}.
\newblock
  \href{http://dx.doi.org/10.1021/acsami.3c08638}{\emph{\bibinfo{journal}{ACS
  Appl. Mater. Interfaces}}} \textbf{\bibinfo{volume}{15}},
  \bibinfo{pages}{49712--49726} (\bibinfo{year}{2023}).

\bibitem{Bornholt2016}
\bibinfo{author}{Bornholt, J.} \emph{et~al.}
\newblock \bibinfo{title}{A {DNA}-based archival storage system}.
\newblock In \emph{\bibinfo{booktitle}{Proceedings of the Twenty-First
  International Conference on Architectural Support for Programming Languages
  and Operating Systems}}, ASPLOS '16, \bibinfo{pages}{637–649}
  (\bibinfo{publisher}{Association for Computing Machinery},
  \bibinfo{address}{New York, NY, USA}, \bibinfo{year}{2016}).

\bibitem{Akram2018}
\bibinfo{author}{Akram, F.}, \bibinfo{author}{ul~Haq, I.},
  \bibinfo{author}{Ali, H.} \& \bibinfo{author}{Laghari, A.~T.}
\newblock \bibinfo{title}{Trends to store digital data in {DNA}: an overview}.
\newblock
  \href{http://dx.doi.org/10.1007/s11033-018-4280-y}{\emph{\bibinfo{journal}{Mol.
  Biol. Rep.}}} \textbf{\bibinfo{volume}{45}}, \bibinfo{pages}{1479–1490}
  (\bibinfo{year}{2018}).

\bibitem{Ceze2019}
\bibinfo{author}{Ceze, L.}, \bibinfo{author}{Nivala, J.} \&
  \bibinfo{author}{Strauss, K.}
\newblock \bibinfo{title}{Molecular digital data storage using {DNA}}.
\newblock
  \href{http://dx.doi.org/10.1038/s41576-019-0125-3}{\emph{\bibinfo{journal}{Nat.
  Rev. Genet.}}} \textbf{\bibinfo{volume}{20}}, \bibinfo{pages}{456–466}
  (\bibinfo{year}{2019}).

\bibitem{Janzakova2021b}
\bibinfo{author}{Janzakova, K.} \emph{et~al.}
\newblock
  \href{http://dx.doi.org/10.6084/m9.figshare.16814710.v1}{\bibinfo{title}{{Dataset
  for: Analog Programing of Conducting Polymer Dendritic Interconnections and
  Control of their Morphology}}} (\bibinfo{year}{2021}).

\bibitem{AkaiKasaya2020}
\bibinfo{author}{Akai-Kasaya, M.} \emph{et~al.}
\newblock \bibinfo{title}{Evolving conductive polymer neural networks on
  wetware}.
\newblock
  \href{http://dx.doi.org/10.35848/1347-4065/ab8e06}{\emph{\bibinfo{journal}{Jpn.
  J. Appl. Phys.}}} \textbf{\bibinfo{volume}{59}}, \bibinfo{pages}{060601}
  (\bibinfo{year}{2020}).

\bibitem{Petrauskas2021}
\bibinfo{author}{Petrauskas, L.} \emph{et~al.}
\newblock \bibinfo{title}{Nonlinear behavior of dendritic polymer networks for
  reservoir computing}.
\newblock
  \href{http://dx.doi.org/10.1002/aelm.202100330}{\emph{\bibinfo{journal}{Adv.
  Electron. Mater.}}} \textbf{\bibinfo{volume}{8}}, \bibinfo{pages}{2100330}
  (\bibinfo{year}{2021}).

\bibitem{Janzakova2023}
\bibinfo{author}{Janzakova, K.} \emph{et~al.}
\newblock \bibinfo{title}{Structural plasticity for neuromorphic networks with
  electropolymerized dendritic pedot connections}.
\newblock
  \href{http://dx.doi.org/10.1038/s41467-023-43887-8}{\emph{\bibinfo{journal}{Nat.
  Commun.}}} \textbf{\bibinfo{volume}{14}}, \bibinfo{pages}{8143}
  (\bibinfo{year}{2023}).

\bibitem{Scholaert2024}
\bibinfo{author}{Scholaert, C.}, \bibinfo{author}{Coffinier, Y.},
  \bibinfo{author}{Pecqueur, S.} \& \bibinfo{author}{Alibart, F.}
\newblock
  \href{http://dx.doi.org/10.48550/arXiv.2407.19847}{\bibinfo{title}{Brain-inspired
  polymer dendrite networks for morphology-dependent computing hardware}}
  (\bibinfo{year}{2024}).

\bibitem{Janzakova2021a}
\bibinfo{author}{Janzakova, K.} \emph{et~al.}
\newblock \bibinfo{title}{Analog programing of conducting-polymer dendritic
  interconnections and control of their morphology}.
\newblock
  \href{http://dx.doi.org/10.1038/s41467-021-27274-9}{\emph{\bibinfo{journal}{Nat.
  Commun.}}} \textbf{\bibinfo{volume}{12}}, \bibinfo{pages}{6898}
  (\bibinfo{year}{2021}).

\bibitem{Scholaert2022}
\bibinfo{author}{Scholaert, C.}, \bibinfo{author}{Janzakova, K.},
  \bibinfo{author}{Coffinier, Y.}, \bibinfo{author}{Alibart, F.} \&
  \bibinfo{author}{Pecqueur, S.}
\newblock \bibinfo{title}{Plasticity of conducting polymer dendrites to bursts
  of voltage spikes in phosphate buffered saline}.
\newblock
  \href{http://dx.doi.org/10.1088/2634-4386/ac9b85}{\emph{\bibinfo{journal}{Neuromorphic
  Comput. Eng.}}} \textbf{\bibinfo{volume}{2}}, \bibinfo{pages}{044010}
  (\bibinfo{year}{2022}).

\bibitem{Baron2024a}
\bibinfo{author}{Baron, A.}, \bibinfo{author}{Hernandez-Balaguera, E.} \&
  \bibinfo{author}{Pecqueur, S.}
\newblock \bibinfo{title}{Correlation between electrochemical relaxations and
  morphologies of conducting polymer dendrites}.
\newblock
  \href{http://dx.doi.org/10.1149/2754-2734/ad9bcb}{\emph{\bibinfo{journal}{ECS
  Adv.}}} \textbf{\bibinfo{volume}{4}}, \bibinfo{pages}{044001}
  (\bibinfo{year}{2024}).

\bibitem{Baron2024}
\bibinfo{author}{Baron, A.}, \bibinfo{author}{Balaguera, E.~H.} \&
  \bibinfo{author}{Pecqueur, S.}
\newblock \bibinfo{title}{A compact electrochemical model for a conducting
  polymer dendrite impedance}.
\newblock In \emph{\bibinfo{booktitle}{2024 International Workshop on Impedance
  Spectroscopy (IWIS)}}, \bibinfo{pages}{1–5} (\bibinfo{year}{2024}).

\bibitem{Baron2025}
\bibinfo{author}{Baron, A.}, \bibinfo{author}{Hernandez-Balaguera, E.},
  \bibinfo{author}{Scholaert, C.}, \bibinfo{author}{Alibart, F.} \&
  \bibinfo{author}{Pecqueur, S.}
\newblock
  \href{http://dx.doi.org/10.48550/arXiv.2504.12861}{\bibinfo{title}{Hardware
  implementation of tunable fractional-order capacitors by morphogenesis of
  conducting polymer dendrites}} (\bibinfo{year}{2025}).

\bibitem{Koizumi2018}
\bibinfo{author}{Koizumi, Y.} \emph{et~al.}
\newblock \bibinfo{title}{Synthesis of
  poly(3,4-ethylenedioxythiophene)–platinum and
  poly(3,4-ethylenedioxythiophene)–poly(styrenesulfonate) hybrid fibers by
  alternating current bipolar electropolymerization}.
\newblock
  \href{http://dx.doi.org/10.1021/acs.langmuir.8b00408}{\emph{\bibinfo{journal}{Langmuir}}}
  \textbf{\bibinfo{volume}{34}}, \bibinfo{pages}{7598–7603}
  (\bibinfo{year}{2018}).

\bibitem{Dang2014}
\bibinfo{author}{Dang, N.~T.}, \bibinfo{author}{Akai-Kasaya, M.},
  \bibinfo{author}{Asai, T.}, \bibinfo{author}{Saito, A.} \&
  \bibinfo{author}{Kuwahara, Y.}
\newblock \bibinfo{title}{Study on electropolymerization micro-wiring system
  imitating axonal growth of artificial neurons towards machine learning}.
\newblock In \emph{\bibinfo{booktitle}{2nd Int. Symp. on the Functionality of
  Organized Nanostructures (FON’14) (Tokyo, Japan 26–28 November 2014)}}
  (\bibinfo{year}{2014}).
\newblock
  \urlprefix\url{https://www.jstage.jst.go.jp/article/pscjspe/2014A/0/2014A_769/_pdf}.

\bibitem{Koizumi2016}
\bibinfo{author}{Koizumi, Y.} \emph{et~al.}
\newblock \bibinfo{title}{Electropolymerization on wireless electrodes towards
  conducting polymer microfibre networks}.
\newblock
  \href{http://dx.doi.org/10.1038/ncomms10404}{\emph{\bibinfo{journal}{Nat.
  Commun.}}} \textbf{\bibinfo{volume}{7}}, \bibinfo{pages}{10404}
  (\bibinfo{year}{2016}).

\bibitem{Ohira2017}
\bibinfo{author}{Ohira, M.}, \bibinfo{author}{Koizumi, Y.},
  \bibinfo{author}{Nishiyama, H.}, \bibinfo{author}{Tomita, I.} \&
  \bibinfo{author}{Inagi, S.}
\newblock \bibinfo{title}{Synthesis of linear {PEDOT} fibers by {AC}-bipolar
  electropolymerization in a micro-space}.
\newblock
  \href{http://dx.doi.org/10.1038/pj.2016.100}{\emph{\bibinfo{journal}{Polym.
  J.}}} \textbf{\bibinfo{volume}{49}}, \bibinfo{pages}{163–167}
  (\bibinfo{year}{2017}).

\bibitem{Watanabe2018}
\bibinfo{author}{Watanabe, T.} \emph{et~al.}
\newblock \bibinfo{title}{In-plane growth of poly(3,4-ethylenedioxythiophene)
  films on a substrate surface by bipolar electropolymerization}.
\newblock
  \href{http://dx.doi.org/10.1021/acsmacrolett.8b00170}{\emph{\bibinfo{journal}{ACS
  Macro Lett.}}} \textbf{\bibinfo{volume}{7}}, \bibinfo{pages}{551–555}
  (\bibinfo{year}{2018}).

\bibitem{Chen2023}
\bibinfo{author}{Chen, Z.} \emph{et~al.}
\newblock \bibinfo{title}{{AC}-bipolar electropolymerization of
  3,4-ethylenedioxythiophene in ionic liquids}.
\newblock
  \href{http://dx.doi.org/10.1021/acs.langmuir.3c00120}{\emph{\bibinfo{journal}{Langmuir}}}
  \textbf{\bibinfo{volume}{39}}, \bibinfo{pages}{4450–4455}
  (\bibinfo{year}{2023}).

\bibitem{Eickenscheidt2019}
\bibinfo{author}{Eickenscheidt, M.}, \bibinfo{author}{Singler, E.} \&
  \bibinfo{author}{Stieglitz, T.}
\newblock \bibinfo{title}{Pulsed electropolymerization of {PEDOT} enabling
  controlled branching}.
\newblock
  \href{http://dx.doi.org/10.1038/s41428-019-0213-4}{\emph{\bibinfo{journal}{Polym.
  J.}}} \textbf{\bibinfo{volume}{51}}, \bibinfo{pages}{1029–1036}
  (\bibinfo{year}{2019}).

\bibitem{Cucchi2021}
\bibinfo{author}{Cucchi, M.} \emph{et~al.}
\newblock \bibinfo{title}{Reservoir computing with biocompatible organic
  electrochemical networks for brain-inspired biosignal classification}.
\newblock
  \href{http://dx.doi.org/10.1126/sciadv.abh0693}{\emph{\bibinfo{journal}{Sci.
  Adv.}}} \textbf{\bibinfo{volume}{7}}, \bibinfo{pages}{eabh0693}
  (\bibinfo{year}{2021}).

\bibitem{Cucchi2021a}
\bibinfo{author}{Cucchi, M.} \emph{et~al.}
\newblock \bibinfo{title}{Directed growth of dendritic polymer networks for
  organic electrochemical transistors and artificial synapses}.
\newblock
  \href{http://dx.doi.org/10.1002/aelm.202100586}{\emph{\bibinfo{journal}{Adv.
  Electron. Mater.}}} \textbf{\bibinfo{volume}{7}}, \bibinfo{pages}{2100586}
  (\bibinfo{year}{2021}).

\bibitem{Hagiwara2021}
\bibinfo{author}{Hagiwara, N.}, \bibinfo{author}{Sekizaki, S.},
  \bibinfo{author}{Kuwahara, Y.}, \bibinfo{author}{Asai, T.} \&
  \bibinfo{author}{Akai-Kasaya, M.}
\newblock \bibinfo{title}{Long- and short-term conductance control of
  artificial polymer wire synapses}.
\newblock
  \href{http://dx.doi.org/10.3390/polym13020312}{\emph{\bibinfo{journal}{Polymers}}}
  \textbf{\bibinfo{volume}{13}}, \bibinfo{pages}{312} (\bibinfo{year}{2021}).

\bibitem{Hagiwara2023}
\bibinfo{author}{Hagiwara, N.}, \bibinfo{author}{Asai, T.},
  \bibinfo{author}{Ando, K.} \& \bibinfo{author}{Akai-Kasaya, M.}
\newblock \bibinfo{title}{Fabrication and training of {3D} conductive polymer
  networks for neuromorphic wetware}.
\newblock
  \href{http://dx.doi.org/10.1002/adfm.202300903}{\emph{\bibinfo{journal}{Adv.
  Funct. Mater.}}} \textbf{\bibinfo{volume}{33}}, \bibinfo{pages}{2300903}
  (\bibinfo{year}{2023}).

\bibitem{Watanabe2024}
\bibinfo{author}{Watanabe, S.} \emph{et~al.}
\newblock \bibinfo{title}{{PEDOT:PSS} wire: A two-terminal synaptic device for
  operation in electrolyte and saline solutions}.
\newblock
  \href{http://dx.doi.org/10.1021/acsami.4c12037}{\emph{\bibinfo{journal}{ACS
  Appl. Mater. Interface}}} \textbf{\bibinfo{volume}{16}},
  \bibinfo{pages}{54636–54644} (\bibinfo{year}{2024}).

\bibitem{Janzakova2021}
\bibinfo{author}{Janzakova, K.} \emph{et~al.}
\newblock \bibinfo{title}{Dendritic organic electrochemical transistors grown
  by electropolymerization for {3D} neuromorphic engineering}.
\newblock
  \href{http://dx.doi.org/10.1002/advs.202102973}{\emph{\bibinfo{journal}{Adv.
  Sci.}}} \textbf{\bibinfo{volume}{8}}, \bibinfo{pages}{2102973}
  (\bibinfo{year}{2021}).

\bibitem{Cutler2005}
\bibinfo{author}{Cutler, C.~A.}, \bibinfo{author}{Bouguettaya, M.},
  \bibinfo{author}{Kang, T.-S.} \& \bibinfo{author}{Reynolds, J.~R.}
\newblock \bibinfo{title}{Alkoxysulfonate-functionalized {PEDOT}
  polyelectrolyte multilayer films: Electrochromic and hole transport
  materials}.
\newblock
  \href{http://dx.doi.org/10.1021/ma047396+}{\emph{\bibinfo{journal}{Macromolecules}}}
  \textbf{\bibinfo{volume}{38}}, \bibinfo{pages}{3068–3074}
  (\bibinfo{year}{2005}).

\bibitem{Ghazal2023}
\bibinfo{author}{Ghazal, M.} \emph{et~al.}
\newblock \bibinfo{title}{Electropolymerization processing of side-chain
  engineered {EDOT} for high performance microelectrode arrays}.
\newblock
  \href{http://dx.doi.org/10.1016/j.bios.2023.115538}{\emph{\bibinfo{journal}{Biosens.
  Bioelectron.}}} \textbf{\bibinfo{volume}{237}}, \bibinfo{pages}{115538}
  (\bibinfo{year}{2023}).

\bibitem{Paulsen2020}
\bibinfo{author}{Paulsen, B.~D.}, \bibinfo{author}{Tybrandt, K.},
  \bibinfo{author}{Stavrinidou, E.} \& \bibinfo{author}{Rivnay, J.}
\newblock \bibinfo{title}{Organic mixed ionic–electronic conductors}.
\newblock
  \href{http://dx.doi.org/10.1038/s41563-019-0435-z}{\emph{\bibinfo{journal}{Nat.
  Mater.}}} \textbf{\bibinfo{volume}{19}}, \bibinfo{pages}{13–26}
  (\bibinfo{year}{2020}).

\bibitem{Tropp2023}
\bibinfo{author}{Tropp, J.}, \bibinfo{author}{Meli, D.} \&
  \bibinfo{author}{Rivnay, J.}
\newblock \bibinfo{title}{Organic mixed conductors for electrochemical
  transistors}.
\newblock
  \href{http://dx.doi.org/10.1016/j.matt.2023.05.001}{\emph{\bibinfo{journal}{Matter}}}
  \textbf{\bibinfo{volume}{6}}, \bibinfo{pages}{3132–3164}
  (\bibinfo{year}{2023}).

\bibitem{Wang2023}
\bibinfo{author}{Wang, X.} \& \bibinfo{author}{Zhao, K.}
\newblock \bibinfo{title}{A continuum theory of organic mixed ionic-electronic
  conductors of phase separation}.
\newblock
  \href{http://dx.doi.org/10.1016/j.jmps.2022.105178}{\emph{\bibinfo{journal}{J.
  Mech. Phys. Solids}}} \textbf{\bibinfo{volume}{172}}, \bibinfo{pages}{105178}
  (\bibinfo{year}{2023}).

\bibitem{Wu2023}
\bibinfo{author}{Wu, R.}, \bibinfo{author}{Paulsen, B.~D.},
  \bibinfo{author}{Ma, Q.}, \bibinfo{author}{McCulloch, I.} \&
  \bibinfo{author}{Rivnay, J.}
\newblock \bibinfo{title}{Quantitative composition and mesoscale ion
  distribution in p-type organic mixed ionic-electronic conductors}.
\newblock
  \href{http://dx.doi.org/10.1021/acsami.3c04449}{\emph{\bibinfo{journal}{ACS
  Appl. Mater. Interfaces}}} \textbf{\bibinfo{volume}{15}},
  \bibinfo{pages}{30553–30566} (\bibinfo{year}{2023}).

\bibitem{Gkoupidenis2024}
\bibinfo{author}{Gkoupidenis, P.} \emph{et~al.}
\newblock \bibinfo{title}{Organic mixed conductors for bioinspired
  electronics}.
\newblock
  \href{http://dx.doi.org/10.1038/s41578-023-00622-5}{\emph{\bibinfo{journal}{Nat.
  Rev. Mater.}}} \textbf{\bibinfo{volume}{9}}, \bibinfo{pages}{134–149}
  (\bibinfo{year}{2024}).

\bibitem{Tommasini2022}
\bibinfo{author}{Tommasini, G.} \emph{et~al.}
\newblock \bibinfo{title}{Seamless integration of bioelectronic interface in an
  animal model via in vivo polymerization of conjugated oligomers}.
\newblock
  \href{http://dx.doi.org/10.1016/j.bioactmat.2021.08.025}{\emph{\bibinfo{journal}{Bioact.
  Mater.}}} \textbf{\bibinfo{volume}{10}}, \bibinfo{pages}{107--116}
  (\bibinfo{year}{2022}).

\bibitem{Stavrinidou2017}
\bibinfo{author}{Stavrinidou, E.} \emph{et~al.}
\newblock \bibinfo{title}{In vivo polymerization and manufacturing of wires and
  supercapacitors in plants}.
\newblock
  \href{http://dx.doi.org/10.1073/pnas.1616456114}{\emph{\bibinfo{journal}{Proc.
  Natl. Acad. Sci.}}} \textbf{\bibinfo{volume}{114}},
  \bibinfo{pages}{2807--2812} (\bibinfo{year}{2017}).

\bibitem{Dufil2020}
\bibinfo{author}{Dufil, G.} \emph{et~al.}
\newblock \bibinfo{title}{Enzyme-assisted in vivo polymerisation of conjugated
  oligomer based conductors}.
\newblock
  \href{http://dx.doi.org/10.1039/D0TB00212G}{\emph{\bibinfo{journal}{J. Mater.
  Chem. B}}} \textbf{\bibinfo{volume}{8}}, \bibinfo{pages}{4221--4227}
  (\bibinfo{year}{2020}).

\bibitem{Priyadarshini2023}
\bibinfo{author}{Priyadarshini, D.} \emph{et~al.}
\newblock \bibinfo{title}{Enzymatically polymerized organic conductors on model
  lipid membranes}.
\newblock
  \href{http://dx.doi.org/10.1021/acs.langmuir.3c00654}{\emph{\bibinfo{journal}{Langmuir}}}
  \textbf{\bibinfo{volume}{39}}, \bibinfo{pages}{8196--8204}
  (\bibinfo{year}{2023}).

\bibitem{Strakosas2023}
\bibinfo{author}{Strakosas, X.} \emph{et~al.}
\newblock \bibinfo{title}{Metabolite-induced in vivo fabrication of
  substrate-free organic bioelectronics}.
\newblock
  \href{http://dx.doi.org/10.1126/science.adc9998}{\emph{\bibinfo{journal}{Science}}}
  \textbf{\bibinfo{volume}{379}}, \bibinfo{pages}{795--802}
  (\bibinfo{year}{2023}).

\bibitem{Fleury_1993}
\bibinfo{author}{Fleury, V.}, \bibinfo{author}{Chazalviel, J.-N.} \&
  \bibinfo{author}{Rosso, M.}
\newblock \bibinfo{title}{Coupling of drift, diffusion, and electroconvection,
  in the vicinity of growing electrodeposits}.
\newblock
  \href{http://dx.doi.org/https://doi.org/10.1103/PhysRevE.48.1279}{\emph{\bibinfo{journal}{Phys.
  Rev. E}}} \textbf{\bibinfo{volume}{48}}, \bibinfo{pages}{1279--1295}
  (\bibinfo{year}{1993}).

\bibitem{Gueye2016}
\bibinfo{author}{Gueye, M.~N.} \emph{et~al.}
\newblock \bibinfo{title}{Structure and dopant engineering in {PEDOT} thin
  films: Practical tools for a dramatic conductivity enhancement}.
\newblock
  \href{http://dx.doi.org/10.1021/acs.chemmater.6b01035}{\emph{\bibinfo{journal}{Chem.
  Mater.}}} \textbf{\bibinfo{volume}{28}}, \bibinfo{pages}{3462–3468}
  (\bibinfo{year}{2016}).

\bibitem{Adamczyk2009}
\bibinfo{author}{Adamczyk, Z.}, \bibinfo{author}{Jachimska, B.},
  \bibinfo{author}{Jasi\'{n}ski, T.}, \bibinfo{author}{Warszy\'{n}ski, P.} \&
  \bibinfo{author}{Wasilewska, M.}
\newblock \bibinfo{title}{Structure of poly (sodium 4-styrenesulfonate) ({PSS})
  in electrolyte solutions: Theoretical modeling and measurements}.
\newblock
  \href{http://dx.doi.org/10.1016/J.COLSURFA.2009.01.035}{\emph{\bibinfo{journal}{Colloids
  Surf. A: Physicochem. Eng. Asp.}}} \textbf{\bibinfo{volume}{343}},
  \bibinfo{pages}{96–103} (\bibinfo{year}{2009}).

\bibitem{LaBarbera2015}
\bibinfo{author}{La~Barbera, S.}, \bibinfo{author}{Vuillaume, D.} \&
  \bibinfo{author}{Alibart, F.}
\newblock \bibinfo{title}{Filamentary switching: Synaptic plasticity through
  device volatility}.
\newblock
  \href{http://dx.doi.org/10.1021/nn506735m}{\emph{\bibinfo{journal}{ACS
  Nano}}} \textbf{\bibinfo{volume}{9}}, \bibinfo{pages}{941–949}
  (\bibinfo{year}{2015}).

\bibitem{Zhang2022}
\bibinfo{author}{Zhang, X.}, \bibinfo{author}{Wu, C.}, \bibinfo{author}{Lv,
  Y.}, \bibinfo{author}{Zhang, Y.} \& \bibinfo{author}{Liu, W.}
\newblock \bibinfo{title}{High-performance flexible polymer memristor based on
  stable filamentary switching}.
\newblock
  \href{http://dx.doi.org/10.1021/acs.nanolett.2c02765}{\emph{\bibinfo{journal}{Nano
  Letters}}} \textbf{\bibinfo{volume}{22}}, \bibinfo{pages}{7246–7253}
  (\bibinfo{year}{2022}).

\bibitem{Cheng2008}
\bibinfo{author}{Cheng, N.-S.}
\newblock \bibinfo{title}{Formula for the viscosity of a glycerol-water
  mixture}.
\newblock
  \href{http://dx.doi.org/10.1021/ie071349z}{\emph{\bibinfo{journal}{Ind. Eng.
  Chem. Res.}}} \textbf{\bibinfo{volume}{47}}, \bibinfo{pages}{3285–3288}
  (\bibinfo{year}{2008}).

\bibitem{Kumar2022}
\bibinfo{author}{Kumar, A.}, \bibinfo{author}{Janzakova, K.},
  \bibinfo{author}{Coffinier, Y.}, \bibinfo{author}{Pecqueur, S.} \&
  \bibinfo{author}{Alibart, F.}
\newblock \bibinfo{title}{Theoretical modeling of dendrite growth from
  conductive wire electro-polymerization}.
\newblock
  \href{http://dx.doi.org/10.1038/s41598-022-10082-6}{\emph{\bibinfo{journal}{Sci.
  Rep.}}} \textbf{\bibinfo{volume}{12}}, \bibinfo{pages}{6395}
  (\bibinfo{year}{2022}).

\bibitem{Sheely1932}
\bibinfo{author}{Sheely, M.~L.}
\newblock \bibinfo{title}{Glycerol viscosity tables}.
\newblock
  \href{http://dx.doi.org/10.1021/ie50273a022}{\emph{\bibinfo{journal}{Ind.
  Eng. Chem. Res.}}} \textbf{\bibinfo{volume}{24}},
  \bibinfo{pages}{1060–1064} (\bibinfo{year}{1932}).

\bibitem{Yanniotis2006}
\bibinfo{author}{Yanniotis, S.}, \bibinfo{author}{Skaltsi, S.} \&
  \bibinfo{author}{Karaburnioti, S.}
\newblock \bibinfo{title}{Effect of moisture content on the viscosity of honey
  at different temperatures}.
\newblock
  \href{http://dx.doi.org/10.1016/j.jfoodeng.2004.12.017}{\emph{\bibinfo{journal}{J.
  Food Eng.}}} \textbf{\bibinfo{volume}{72}}, \bibinfo{pages}{372–377}
  (\bibinfo{year}{2006}).

\bibitem{Akoudad2000}
\bibinfo{author}{Akoudad, S.} \& \bibinfo{author}{Roncali, J.}
\newblock \bibinfo{title}{Modification of the electrochemical and electronic
  properties of electrogenerated poly(3,4-ethylenedioxythiophene) by
  hydroxymethyl and oligo(oxyethylene) substituents}.
\newblock
  \href{http://dx.doi.org/10.1016/S1388-2481(99)00147-2}{\emph{\bibinfo{journal}{Electrochem.
  Commun.}}} \textbf{\bibinfo{volume}{2}}, \bibinfo{pages}{72–76}
  (\bibinfo{year}{2000}).

\bibitem{Perepichka2002}
\bibinfo{author}{Perepichka, I.~F.}, \bibinfo{author}{Besbes, M.},
  \bibinfo{author}{Levillain, E.}, \bibinfo{author}{Sall\'e, M.} \&
  \bibinfo{author}{Roncali, J.}
\newblock \bibinfo{title}{Hydrophilic oligo(oxyethylene)-derivatized
  poly(3,4-ethylenedioxythiophenes): Cation-responsive
  optoelectroelectrochemical properties and solid-state chromism}.
\newblock
  \href{http://dx.doi.org/10.1021/cm010756c}{\emph{\bibinfo{journal}{Chem.
  Mater.}}} \textbf{\bibinfo{volume}{14}}, \bibinfo{pages}{449–457}
  (\bibinfo{year}{2002}).

\bibitem{Stavrinidou2014}
\bibinfo{author}{Stavrinidou, E.} \emph{et~al.}
\newblock \bibinfo{title}{Engineering hydrophilic conducting composites with
  enhanced ion mobility}.
\newblock
  \href{http://dx.doi.org/10.1039/C3CP54061H}{\emph{\bibinfo{journal}{Phys.
  Chem. Chem. Phys.}}} \textbf{\bibinfo{volume}{16}},
  \bibinfo{pages}{2275–2279} (\bibinfo{year}{2014}).

\bibitem{Mantione2020}
\bibinfo{author}{Mantione, D.} \emph{et~al.}
\newblock \bibinfo{title}{Thiophene-based trimers for in vivo electronic
  functionalization of tissues}.
\newblock
  \href{http://dx.doi.org/10.1021/acsaelm.0c00861}{\emph{\bibinfo{journal}{ACS
  Appl. Electron. Materials}}} \textbf{\bibinfo{volume}{2}},
  \bibinfo{pages}{4065–4071} (\bibinfo{year}{2020}).

\bibitem{Schweiss2005}
\bibinfo{author}{Schweiss, R.}, \bibinfo{author}{L\"ubben, J.~F.},
  \bibinfo{author}{Johannsmann, D.} \& \bibinfo{author}{Knoll, W.}
\newblock \bibinfo{title}{Electropolymerization of ethylene dioxythiophene
  ({EDOT}) in micellar aqueous solutions studied by electrochemical quartz
  crystal microbalance and surface plasmon resonance}.
\newblock
  \href{http://dx.doi.org/10.1016/j.electacta.2004.11.032}{\emph{\bibinfo{journal}{Electrochim.
  Acta}}} \textbf{\bibinfo{volume}{50}}, \bibinfo{pages}{2849–2856}
  (\bibinfo{year}{2005}).

\bibitem{Stephan1998}
\bibinfo{author}{St\'ephan, O.} \emph{et~al.}
\newblock \bibinfo{title}{Electrochemical behaviour of
  3,4-ethylenedioxythiophene functionalized by a sulphonate group. application
  to the preparation of poly(3, 4-ethylenedioxythiophene) having permanent
  cation-exchange properties}.
\newblock
  \href{http://dx.doi.org/10.1016/S0022-0728(97)00548-2}{\emph{\bibinfo{journal}{J.
  Electroanal. Chem.}}} \textbf{\bibinfo{volume}{443}},
  \bibinfo{pages}{217–226} (\bibinfo{year}{1998}).

\bibitem{Yano2019}
\bibinfo{author}{Yano, H.}, \bibinfo{author}{Kudo, K.},
  \bibinfo{author}{Marumo, K.} \& \bibinfo{author}{Okuzaki, H.}
\newblock \bibinfo{title}{Fully soluble self-doped
  poly(3,4-ethylenedioxythiophene) with an electrical conductivity greater than
  1000 {S} cm$^{-1}$}.
\newblock
  \href{http://dx.doi.org/10.1126/sciadv.aav9492}{\emph{\bibinfo{journal}{Sci.
  Adv.}}} \textbf{\bibinfo{volume}{5}}, \bibinfo{pages}{eaav9492}
  (\bibinfo{year}{2019}).

\bibitem{Wan2022}
\bibinfo{author}{Wan, Y.}, \bibinfo{author}{Zhang, X.}, \bibinfo{author}{Bazan,
  G.~C.}, \bibinfo{author}{Nguyen, T.-Q.} \& \bibinfo{author}{Lu, G.}
\newblock \bibinfo{title}{Polarons in conjugated polyelectrolytes: A
  first-principles perspective}.
\newblock
  \href{http://dx.doi.org/10.1002/adfm.202209394}{\emph{\bibinfo{journal}{Adv.
  Funct. Mater.}}} \textbf{\bibinfo{volume}{32}}, \bibinfo{pages}{2209394}
  (\bibinfo{year}{2022}).

\bibitem{Danielsen2022}
\bibinfo{author}{Danielsen, S. P.~O.} \emph{et~al.}
\newblock \bibinfo{title}{Ionic tunability of conjugated polyelectrolyte
  solutions}.
\newblock
  \href{http://dx.doi.org/10.1021/acs.macromol.2c00178}{\emph{\bibinfo{journal}{Macromolecules}}}
  \textbf{\bibinfo{volume}{55}}, \bibinfo{pages}{3437–3448}
  (\bibinfo{year}{2022}).

\bibitem{Chae2024}
\bibinfo{author}{Chae, S.} \emph{et~al.}
\newblock \bibinfo{title}{Impact of molecular weight on the ionic and
  electronic transport of self-doped conjugated polyelectrolytes relevant to
  organic electrochemical transistors}.
\newblock
  \href{http://dx.doi.org/10.1002/adfm.202310852}{\emph{\bibinfo{journal}{Adv.
  Funct. Mater.}}} \textbf{\bibinfo{volume}{34}}, \bibinfo{pages}{2310852}
  (\bibinfo{year}{2024}).

\bibitem{Zeglio2015}
\bibinfo{author}{Zeglio, E.} \emph{et~al.}
\newblock \bibinfo{title}{Conjugated polyelectrolyte blends for electrochromic
  and electrochemical transistor devices}.
\newblock
  \href{http://dx.doi.org/10.1021/acs.chemmater.5b02501}{\emph{\bibinfo{journal}{Chem.
  Mater.}}} \textbf{\bibinfo{volume}{27}}, \bibinfo{pages}{6385–6393}
  (\bibinfo{year}{2015}).

\bibitem{Zeglio2016}
\bibinfo{author}{Zeglio, E.}
\newblock \emph{\bibinfo{title}{Self-doped Conjugated Polyelectrolytes for
  Bioelectronics Applications}}.
\newblock Ph.D. thesis, \bibinfo{school}{Link\"oping University, Sweden}
  (\bibinfo{year}{2016}).
\newblock
  \urlprefix\url{https://liu.diva-portal.org/smash/get/diva2:1048490/FULLTEXT01.pdf}.

\bibitem{Zeglio2017}
\bibinfo{author}{Zeglio, E.}, \bibinfo{author}{Eriksson, J.},
  \bibinfo{author}{Gabrielsson, R.}, \bibinfo{author}{Solin, N.} \&
  \bibinfo{author}{Ingan\"as, O.}
\newblock \bibinfo{title}{Highly stable conjugated polyelectrolytes for
  water-based hybrid mode electrochemical transistors}.
\newblock
  \href{http://dx.doi.org/10.1002/adma.201605787}{\emph{\bibinfo{journal}{Adv.
  Mater.}}} \textbf{\bibinfo{volume}{29}}, \bibinfo{pages}{1605787}
  (\bibinfo{year}{2017}).

\bibitem{NguyenDang2022}
\bibinfo{author}{Nguyen-Dang, T.} \emph{et~al.}
\newblock \bibinfo{title}{Dual-mode organic electrochemical transistors based
  on self-doped conjugated polyelectrolytes for reconfigurable electronics}.
\newblock
  \href{http://dx.doi.org/10.1002/adma.202200274}{\emph{\bibinfo{journal}{Adv.
  Mater.}}} \textbf{\bibinfo{volume}{34}}, \bibinfo{pages}{2200274}
  (\bibinfo{year}{2022}).

\bibitem{Llanes2023}
\bibinfo{author}{Llanes, L.~C.} \emph{et~al.}
\newblock \bibinfo{title}{Side-chain engineering of self-doped conjugated
  polyelectrolytes for organic electrochemical transistors}.
\newblock
  \href{http://dx.doi.org/10.1039/D3TC00355H}{\emph{\bibinfo{journal}{J. Mater.
  Chem. C}}} \textbf{\bibinfo{volume}{11}}, \bibinfo{pages}{8274–8283}
  (\bibinfo{year}{2023}).

\end{thebibliography}

\end{document}